\documentclass[times,twocolumn,final]{elsarticle}

\usepackage{color}
\usepackage{framed,multirow}
\usepackage{amssymb}
\usepackage{amsmath}
\usepackage{latexsym}
\usepackage{array}
\usepackage{graphics}
\usepackage{subfigure}
\usepackage{multirow}
\usepackage{booktabs}
\usepackage[switch]{lineno}
\usepackage{url}
\usepackage[colorlinks=green]{hyperref}
\journal{Medical Image Analysis}

\begin{document}

% \verso{Junyi Qiu \textit{et~al.}}

\begin{frontmatter}

\title{MyoPS-Net: Myocardial Pathology Segmentation with Flexible Combination of Multi-Sequence CMR Images}

\author[1]{Junyi Qiu}
\author[2]{Lei Li}
\author[1]{Sihan Wang}
\author[1]{Ke Zhang}
\author[3,4]{Yinyin Chen}
\author[3,4]{Shan Yang}
\author[1]{Xiahai Zhuang\corref{cor1}}
\cortext[cor1]{Corresponding author (www.sdspeople.fudan.edu.cn/zhuangxiahai/)}

\address[1]{School of Data Science, Fudan University, Shanghai, China}
\address[2]{Institute of Biomedical Engineering, University of Oxford, Oxford, UK}
\address[3]{Department of Radiology, Zhongshan Hospital, Fudan University, Shanghai, China}
\address[4]{Department of Medical Imaging, Shanghai Medical School, Fudan University and Shanghai Institute of Medical Imaging, Shanghai, China}

% \received{1 May 2013}
% \finalform{10 May 2013}
% \accepted{13 May 2013}
% \availableonline{15 May 2013}
% \communicated{S. Sarkar}

\begin{abstract}
Myocardial pathology segmentation (MyoPS) can be a prerequisite for the accurate diagnosis and treatment planning of myocardial infarction. However, achieving this segmentation is challenging, mainly due to the inadequate and indistinct information from an image.
In this work, we develop an end-to-end deep neural network, referred to as MyoPS-Net, to flexibly combine five-sequence cardiac magnetic resonance (CMR) images for MyoPS.
To extract precise and adequate information, we design an effective yet flexible architecture to extract and fuse cross-modal features.
This architecture can tackle different numbers of CMR images and complex combinations of modalities, with output branches targeting specific pathologies.  
To impose anatomical knowledge on the segmentation results, we first propose a module to regularize myocardium consistency and localize the pathologies, and then introduce an inclusiveness loss to utilize relations between myocardial scars and edema.
We evaluated the proposed MyoPS-Net on two datasets, \textit{i.e.,}
a private one consisting of 50 paired multi-sequence CMR images and
a public one from MICCAI2020 MyoPS Challenge.
Experimental results showed that MyoPS-Net could achieve state-of-the-art performance in various scenarios.
Note that in practical clinics, the subjects may not have full sequences, such as missing LGE CMR or mapping CMR scans.
We therefore conducted extensive experiments to investigate the performance of the proposed method in dealing with such complex combinations of different CMR sequences. 
Results proved the superiority and generalizability of MyoPS-Net, and more importantly, indicated a practical clinical application.
\end{abstract}


\end{frontmatter}

% \linenumbers

\section{Introduction} %%%%% 1 Introduction %%%%%
\begin{figure*}[!t]
\centering
\includegraphics[width=0.9\textwidth]{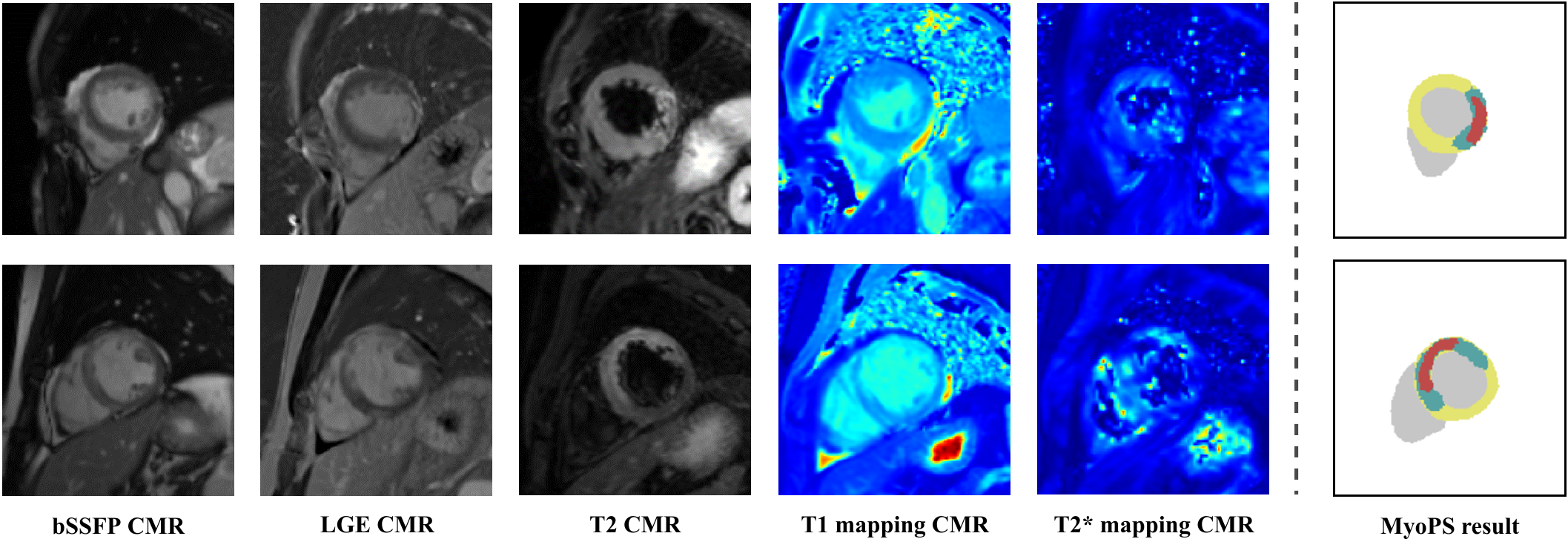}\\[-2ex]
\caption{Examples of the five-sequence CMR images, namely bSSFP, LGE, T2, T1 mapping and T2* mapping CMR. The left five columns display CMR images, and the sixth column presents the MyoPS results overlaid on the ventricles. Scar and edema regions are rendered in dark red and dark green, respectively. Note that here T1 mapping and T2* mapping CMR are displayed in colored images using jet colormap, while in the other figures of this manuscript they are displayed in gray images.}
\label{example}
\end{figure*}

Myocardial infarction (MI), a common cardiovascular disease, is one of the leading causes of death in the world \citep{thygesen2007universal}. 
Quantitative evaluation of myocardial scars and edema is in great demand for effective diagnosis and treatment of MI \citep{richardson2015physiological}. 
In real clinical scenarios, multi-sequence cardiac magnetic resonance (CMR) imaging is usually employed as an efficacious tool for MI evaluation. 
In this work, we propose to consider five clinically available CMR sequences, including balanced steady-state free precession (bSSFP) sequence, late gadolinium enhancement (LGE) sequence, T2-weighted sequence, T1 mapping sequence and T2* mapping sequence. 
Examples of these sequences are presented in Fig.~\ref{example}. 
Specifically, bSSFP CMR can provide clear boundaries of the myocardium and left ventricle (LV) for scar and edema evaluation; 
LGE CMR reveals the scar regions with distinctive brightness compared to the healthy myocardium, while T2, T1 mapping and T2* mapping images provide pathological information without injection of contrast agents. 
To conduct quantitative analysis of multi-sequence CMR images, accurate and efficient segmentation of myocardial pathology regions becomes a necessity. 
Because of the difficulty in manual delineation, which is time-consuming and sometimes subjective, automating this segmentation is highly demanded. 
However, achieving such automatic segmentation is still arduous, mainly due to inferior image quality, blurry pathology boundaries, various pathology appearances and the difficulty in fusing multi-sequence information.

Limited studies have been reported on automatic myocardial scar and edema segmentation. 
A large number of approaches on MyoPS solely focus on one type of pathologies, \textit{i.e.,} either scars or edema. 
For instance, \citet{sandfort2017automatic} and \citet{kolipaka2005segmentation} proposed threshold-based methods to perform scar segmentation, while \citet{kadir2010automatic} developed similar method for edema segmentation. 
Besides, given the difficulty in mining multi-sequence information, most of the reported algorithms barely employed a single sequence for MyoPS, especially LGE CMR which probably contains the most information \citep{MRI}, in spite of the drawback of injecting contrast agents. 
For example, \citet{brahim20213d} proposed a 3D network for myocardial disease segmentation with LGE CMR images. 
However, multi-sequence CMR images can offer supplementary pathology-related features for scar and edema segmentation. 
Effectively extracting instructive features from multi-sequence CMR images becomes essential for MyoPS. 
Furthermore, to better localize the tiny pathologies, existing methods were usually designed in a two-stage manner \citep{myopschallenge-twostage-2,myopschallenge-twostage-1,myopschallenge-twostage-3}, namely first extracting the myocardium and then segmenting scars and edema on the extracted myocardium. 

In this work, we propose an end-to-end deep learning framework for MyoPS combining multi-sequence CMR images, which can be abbreviated as MyoPS-Net. 
The CMR images considered here include the end-diastolic (ED) phase of the cine bSSFP sequence, referred to as C0 image, and four other CMR sequences, \textit{i.e.,} LGE, T2, T1 mapping and T2* mapping, at the same ED phase.
The former provides clear anatomical information, while the latter four offer critical information for the myocardial pathologies, namely scars and edema. 
To deal with the complex information from multi-sequence images, we introduce a cross-modal feature fusion architecture to learn both inter- and intra-sequence pathological features. 
To efficiently localize the pathology regions, we adopt a myocardium prior and consistency module to impose constraints on the myocardium. 
Finally, considering the spatial relations between the pathologies, \textit{i.e.,} scars lying inside edema, we further develop an inclusiveness loss, as a regularization term, to incorporate this prior knowledge into the process of feature learning and segmentation inference. 

The remaining paper is organized as follows: Section~\ref{section2} reviews the related research. The detailed structure of MyoPS-Net is described in Section~\ref{section3}. Section~\ref{section4} illustrates the experiments and results and Section~\ref{section5} presents the discussion and conclusion.

\section{Related work} %%%%% 2 Related work %%%%%
\label{section2}
\begin{figure*}[!t]
\centering
\includegraphics[width=0.99\textwidth]{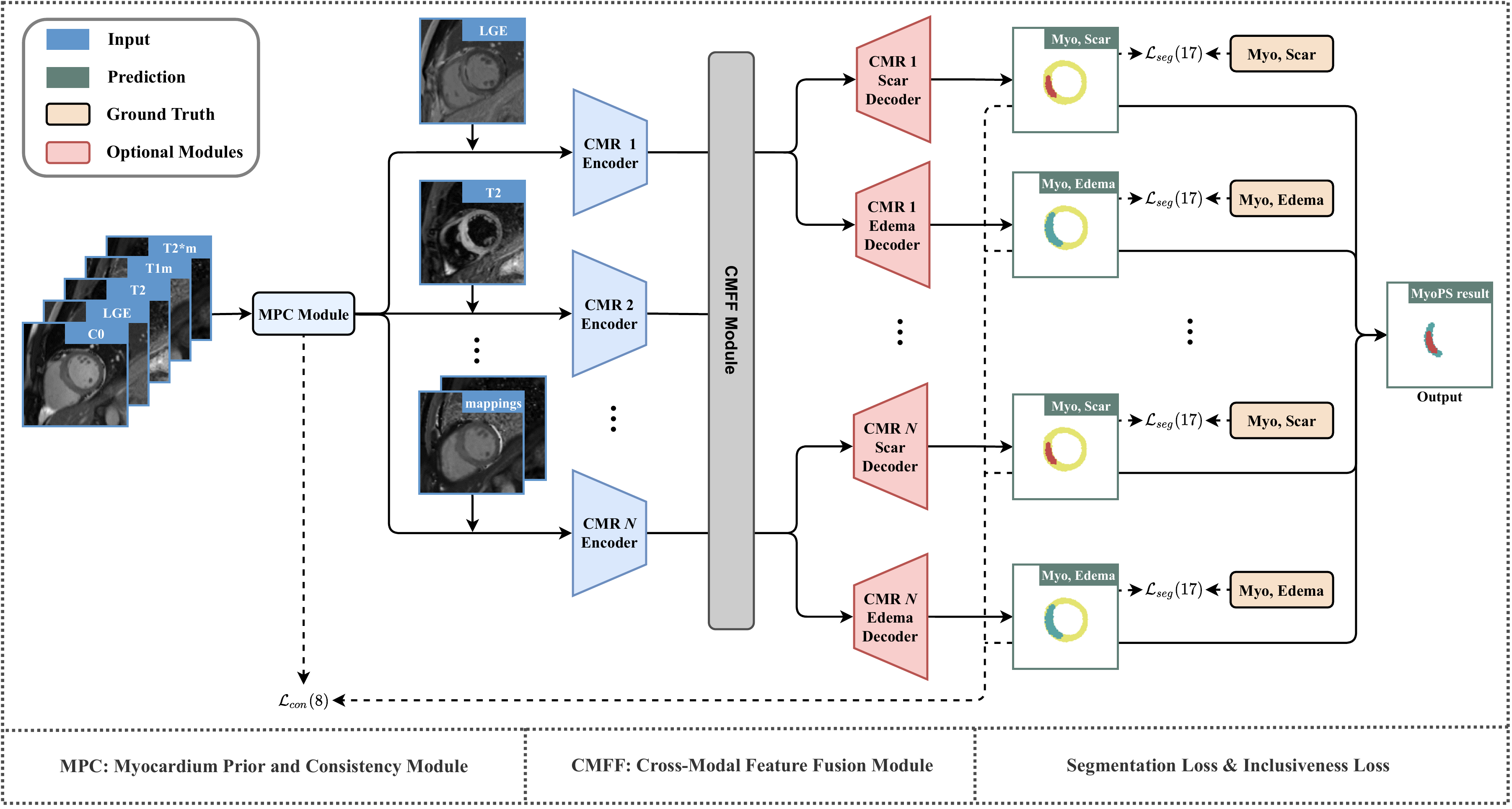}
\caption{The proposed MyoPS-Net consists of three major parts. MPC represents the myocardium prior and consistency module and CMFF denotes the cross-modal feature fusion module. For the optional scar/edema decoders, which are detailed in Section~\ref{3.1}, one can either use one decoder or two decoders to perform pathology segmentation. The parenthesis and numbers are indices of equations in the text, and Myo and LV represent the myocardium and left ventricle, respectively. Here, mappings (T1 mapping and T2* mapping CMR images) are fed into the CMR encoder in two channels.}
\label{structure}
\end{figure*}

\subsection{Myocardial pathology segmentation}
For myocardial pathology segmentation, there exists a series of conventional algorithms. 
For example, \citet{kolipaka2005segmentation} and \citet{sandfort2017automatic} utilized threshold methods for scar segmentation based on the intensity differences between healthy and diseased myocardium. 
\citet{kadir2010automatic} combined the threshold segmentation with a morphological filtering algorithm for edema segmentation. Considering the drawbacks of thresholding methods, \citet{ukwatta2015myocardial} formulated a continuous max-flow optimization problem for edema segmentation. 
Moreover, graph-cuts were also employed for pathology segmentation \citep{lu2012automated}.

Recently, deep learning (DL)-based algorithms have presented satisfying performance in the field of medical image segmentation. 
In particular, convolution neural networks (CNNs) \citep{albawi2017understanding,sultana2020evolution} have the widest applications in the field of image segmentation. 
For example, \citet{li2020atrial} proposed a LearnGC framework for scar quantification from LGE CMR images by combining CNNs and graph-cuts.
\citet{moccia2019development} introduced fully convolutional networks to quantitatively evaluate myocardial scars. 
Specially, with the emergence of U-Net \citep{unet}, an extremely huge proportion of medical image segmentation frameworks are proposed on the basis of U-Net. 
For instance, \citet{li2020joint,li2022atrialjsqnet} proposed a multi-branch U-Net to perform multi-task learning of joint left atrial segmentation and scar quantification on LGE CMR images. 
\citet{zabihollahy2020fully} performed automatic scar segmentation with a cascaded multi-planar U-Net.

To the best of our knowledge, most of the work merely concentrated on either scar or edema segmentation and usually employed a single CMR sequence.
With the growth of the modeling ability of deep learning frameworks, studies have recently focused on combining multi-sequence information for MyoPS to leverage abundant pathology-related features provided by multi-sequence CMR images. 
A two-stage segmentation strategy has been proposed to segment scars and edema simultaneously \citep{myopschallenge-twostage-2,myopschallenge-twostage-1, myopschallenge-twostage-3}, in a coarse-to-fine manner. \citet{myopschallenge-twostage-3} employed a modified nnUNet \citep{nnunet} to effectively modeling pathological information. 
\citet{CMS-UNet} employed one shared encoder to collect features from multi-sequence CMR images. 
\citet{ApplicationtoCMR} proposed a fully automatic pathology segmentation framework consisting of an anatomical structure segmentation network and a pathological region segmentation network. 
\citet{Dual-Path} introduced a multi-layer fusion to merge features from multi-sequence CMR images. 
From a new perspective, \citet{AWSnet} combinedly utilized deep supervision and deep reinforcement learning for MyoPS. 
Besides, considering the fuzzy boundaries and the class-imbalance problem of scars and edema, \citet{EfficientSeg} proposed a boundary loss to perform MyoPS based on an efficient net \citep{tan2019efficientnet}.

\subsection{Deep learning based multi-modality medical image segmentation}
Multi-modality is a typical characteristic in the field of medical image analysis. 
As multi-modality images could usually provide much more information than single modality ones, studying multi-modality images has thus become a hot spot \citep{journal/MedIA_review/li2022}. 
Recently, some challenges were held for multi-modality medical image segmentation, such as BraTS Challenge \citep{menze2014multimodal}, CHAOS Challenge \citep{chaos}, MM-WHS Challenge \citep{journal/MedIA/zhuang2019}, MS-CMRSeg Challenge \citep{journal/MedIA/zhuang2022} and MyoPS Challenge \citep{myopsbenchmark}. 
These challenges further arouse the research interest towards fusing multi-modality information \citep{zhou2019review}. 
Specifically, the fusion strategy of network architectures can be divided into three categories, namely input-level fusion, layer-level fusion and decision-level fusion. 
The proposed MyoPS-Net introduces a fusion strategy from the layer level.

As for the input-level fusion strategy, medical images from multi-modality are directly concatenated to generate the multi-channel inputs of the network \citep{input-1,input-2,input-3}. 
This strategy obviously treats multi-modality images equally and thus concentrates on the formulation of the network architecture. 
For example, \citet{yang2018automatic} employed a residual U-Net and generative adversarial networks to perform brain tumor segmentation simply by fusing multi-modality images as multi-channel inputs. 
When it comes to the layer-level fusion method, each modality image is individually transferred into the network to extract features. 
Different from input-level strategy, layer-level fusion focuses on modifying the network architectures. 
The multi-modality features are fused inside the layer of the networks and are then synthesized into a unified feature representation for the final segmentation. 
\citet{layer-1} proposed a convolution neural framework based on DenseNets with dense connections between multi-modality features. \citet{Dual-Path} introduced a dual-path strategy to fuse features from multi-sequence CMR images for MyoPS. 
\citet{zhao2021united} combined layer-level fusion with adversarial learning for liver tumor segmentation. 
Further, some disentanglement-based methods were recently introduced for multi-modality segmentation \citep{chen2019robust,pei2021disentangle}. 
In the decision-level strategy, similar to the above layer-level one, multi-modality images are trained with independent networks to generate the corresponding features. 
However, there exist none of feature transmissions between different modalities. 
The multi-modality information is fused at the lower layer of the encoders and these mixed features will be employed to acquire the segmentation results \citep{decision-1}.

\section{Method} %%%%% 3 Method %%%%%
\label{section3}
Fig.~\ref{structure} provides an overview of the proposed MyoPS-Net that can combine different CMR images, such as C0, LGE, T2, T1 mapping and T2* mapping CMR images for MyoPS.
MyoPS-Net is an end-to-end architecture consisting of a cross-modal feature fusion module (Section~\ref{3.1}), two assisting modules for imposing myocardium prior and consistency (Section~\ref{3.2}) and pathology inclusiveness constraints (Section~\ref{3.3}). 
Besides, the proposed MyoPS-Net can be applied to four different scenarios in practical clinics (Section~\ref{3.5}).

\subsection{Cross-modal feature fusion for multi-sequence CMR}
\label{3.1}
Each of the multi-sequence CMR images has a specific influence on the final results of MyoPS. 
For example, LGE CMR is promising in discriminating myocardial scars, but it also provides complementary information for edema segmentation. 
Therefore, it is crucial to fuse multi-sequence features to excavate hidden inter-sequence relations.

\begin{figure}[!t]
\centering
\includegraphics[width=0.99\linewidth]{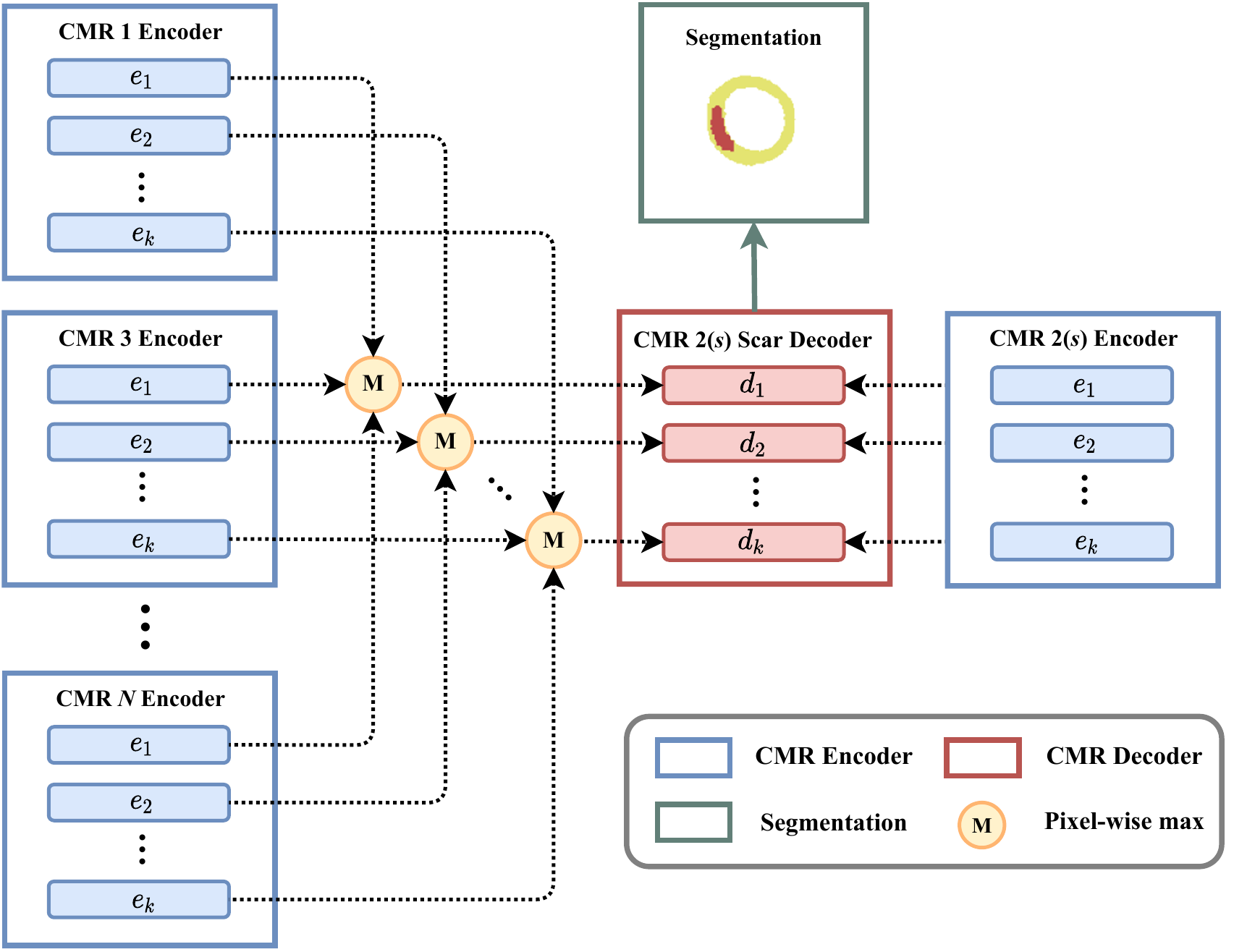}
\caption{Architecture of cross-modal feature fusion module on $s$-th CMR sequence. Here, we take the CMR 2 Scar Decoder ($s=2$) as an example for illustration.}
\label{crossmodal}
\end{figure}

We propose a cross-modal feature fusion (CMFF) module to fuse multi-sequence information. 
CMFF is the main architecture of MyoPS-Net, as Fig.~\ref{structure} shows, and is designed to effectively extract the features contained in the CMR images (via CMR encoders), and fuse them for pathology segmentation (via scar/ edema decoders). 
For feature fusion, 
CMFF adopts two operations, namely max operation and skip connection.
Fig.~\ref{crossmodal} illustrates the detail of network architecture.
Let $\phi_{n}^{l}$ be the multi-scale features of CMR encoders, where $n \in \{1,\cdots,N\}$ denotes the index of CMR sequences except C0, and $l \in \{1,\cdots,k\}$ means the index of layers of encoders.
Therefore, a pixel-wise max operation is performed on these features as follows,
\begin{align}
    \hat{\phi}_{n}^{l} &= \max\{\phi_{1}^{l}, \phi_{2}^{l}, \cdots,\phi_{n-1}^{l}, \phi_{n+1}^{l}, \cdots, \phi_{N}^{l}\},
\end{align}
where $\hat{\phi}_{n}^{l}\ (n \in \{1,2,\cdots,N\})$ means the fused features from all other $N-1$ sequences except the sequence $n$. 
Denote $D_\text{scar}$ and $D_\text{edema}$ as the sets of scar decoders and edema decoders, respectively. 
When the fused features are obtained, we then introduce connections from $\hat{\phi}_{n}^{l}$ to the corresponding layers of pathology (scar or edema) decoder $n$ in the set $D_\text{scar}$ or $D_\text{edema}$. 
With this module, the pathology segmentation networks can learn cross-modal information through multi-scale features from all other sequences. 

In our application, we have four CMR sequences to perform pathology segmentation, \textit{i.e.,} LGE, T2, T1 mapping and T2* mapping CMR. We adopt three encoders for these four CMR sequences, \textit{i.e.,} one for LGE, another for T2 and a third for mappings (T1 mapping and T2* mapping), to extract pathological features. 
For decoders, note that for each CMR sequence above, one can either use two decoders, respectively for scars and edema, to perform pathology segmentation, or use just one to more effectively segment one type of pathology.  
For example, in this work we propose to couple a scar decoder to LGE CMR, an edema decoder to T2 CMR, and a scar decoder to mapping CMR according to the clinical knowledge and experimental results from Section~\ref{4.3}.

\subsection{Constraint for myocardium prior and consistency}
\label{3.2}
Since scars and edema lie in the myocardium, it is straightforward to consider the myocardium as a prior. Existing methods usually segment the myocardium via a separated network and then employ the cropped ROIs for further MyoPS. 
This requires the myocardial segmentation tool to be reliable, since it is the prerequisite for following MyoPS. 
Here, we instead propose to embed a myocardium prior and consistency (MPC) module in the MyoPS-Net to localize the pathologies.

Denote the five sequences of CMR images, namely C0, LGE, T2, T1 mapping and T2* mapping CMR, as $I_{\text{C0}}$, $I_{\text{LGE}}$, $I_{\text{T2}}$, $I_{\text{T1m}}$ and $I_{\text{T2*m}}$. 
The input of the MPC module, as shown in Fig.~\ref{structure}, is obtained by directly concatenating these five-sequence images, which can be explicitly expressed as $x_{\text{MPC}} = \{I_{\text{C0}}, I_{\text{LGE}}, I_{\text{T2}}, I_{\text{T1m}}, I_{\text{T2*m}}\}$. 
Therefore, the MPC module which employs U-Net \citep{unet} as the backbone can merge distinct features of cardiac structure from the five images and achieve the probability map containing the myocardium (Myo) and left ventricle (LV) information, which is represented as $\psi_{\text{MPC}}$.
One can easily concatenate these features with the raw images $I_{\text{LGE}}$, $I_{\text{T2}}$, $I_{\text{T1m}}$ and $I_{\text{T2*m}}$,
\begin{align}
    x_{\text{LGE}} &= \{I_{\text{LGE}}, \psi_{\text{MPC}}\},\\
    x_{\text{T2}} &= \{I_{\text{T2}}, \psi_{\text{MPC}}\},\\
    x_{\text{mappings}} &= \{I_{\text{T1m}}, I_{\text{T2*m}}, \psi_{\text{MPC}}\},
\end{align}
as inputs of pathology networks, which then benefit from the myocardium prior for localizing and segmenting pathologies within the myocardium.  

Therefore, we propose a consistency loss on the MPC module, to regularize the invariability of the myocardium. 
Specifically, we reformulated the probability maps into two parts, \textit{i.e.,} $myo$ and $\overline{myo}$. Here, $myo$ illustrated the probability of being identified as the myocardium, while $\overline{myo}$ indicates the probability of all other structures except the myocardium.
The reformulated probability maps are defined as follows,
\begin{align}
    \hat{\psi}_{\text{MPC}} &= \{\psi_{\text{MPC}}^{myo},\psi_{\text{MPC}}^{\overline{myo}}\},\\
    \hat{\psi}_{s} &= \{\psi_{s}^{myo},\psi_{s}^{\overline{myo}}\},\\
    \hat{\psi}_{e} &= \{\psi_{e}^{myo},\psi_{e}^{\overline{myo}}\},
\end{align}
where $\hat{\psi}_{\text{MPC}}$ and $\psi_{\text{MPC}}$ represent the modified and original probability maps of the MPC module, respectively.
$\hat{\psi}_{s}$ and $\psi_{s}$ are the probability maps where $s \in D_\text{scar}$, while $\hat{\psi}_{e}$ and $\psi_{e}$ stand for the probability maps where $e \in D_\text{edema}$.
The consistency loss is then given by,
\begin{align}
    \mathcal{L}_{con} = \sum_{i\in D_\text{scar} \cup D_\text{edema}} \mathcal{L}_{cos}(\hat{\psi}_{\text{MPC}},\hat{\psi}_{i}),
\end{align}
where
\begin{align}
    \mathcal{L}_{cos}(\hat{\psi}_1,\hat{\psi}_2) = 1 - \frac{\hat{\psi}_1 \cdot \hat{\psi}_2}{\Vert \hat{\psi}_1 \Vert \cdot \Vert \hat{\psi}_2 \Vert},
\end{align}
is the cosine similarity loss and $\Vert \cdot \Vert$ is $L_2$ norm.

\subsection{Constraint for pathology inclusiveness}
\label{3.3}
\begin{figure}[!t]
    \centering
    \includegraphics[width=0.85\linewidth]{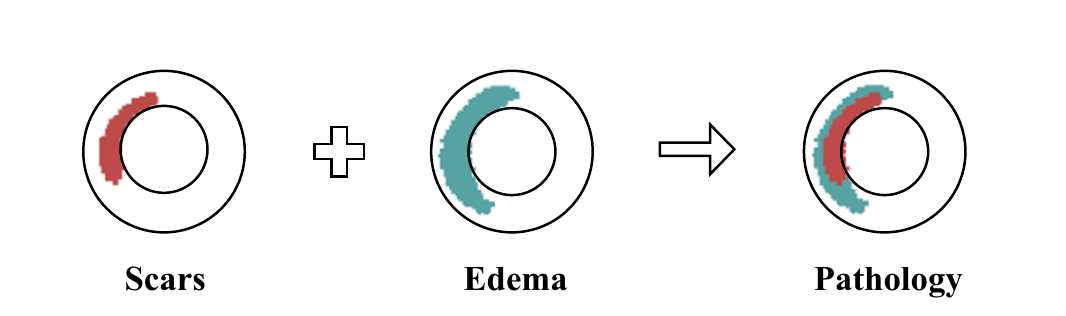}\\[-1ex]
    \caption{The pathology inclusiveness between myocardial scars and edema.}
    \label{inclusive}
\end{figure}

As shown in Fig.~\ref{inclusive}, scars lie inside of edema. 
To embed this prior knowledge, we introduce a pathology inclusiveness (PI) loss in the MyoPS-Net which can be applied to both labeled and unlabeled data.

\subsubsection{Pathology inclusiveness loss for labeled data}
Let $y_\text{scar}$ and $y_\text{edema}$ represent the ground truth of scars and edema, respectively. 
The inclusiveness loss on labeled data ($L$) is then formulated as follows,
\begin{align}
    \label{eq:incloss:L}
    \mathcal{L}_{inc,S}^{L} & = - \frac{1}{\omega_{\text{edema}}} \sum_{s \in D_\text{scar}} \sum_{i \in \Omega} (1-y_\text{edema})\log (1-\psi_{s}^\text{scar}), \\
    \mathcal{L}_{inc,E}^{L} & = - \frac{1}{\omega_{\text{scar}}} \sum_{e \in D_\text{edema}} \sum_{i \in \Omega} y_\text{scar}\log \psi_{e}^\text{edema},
\end{align}
where $\mathcal{L}_{inc,S}^{L}$ and $\mathcal{L}_{inc,E}^{L}$ are respectively for scar and edema decoders, and the superscripts scar and edema refer to the corresponding channel of the probability maps. 
These two factors $\omega_{\text{scar}}, \omega_{\text{edema}}$ are employed to guarantee the robustness of the inclusiveness loss against diverse sizes of pathology areas, with the expressions of $\omega_{\text{scar}} = \sum_y y_\text{scar}, \omega_{\text{edema}} = \sum_y (1-y_\text{edema})$. 
Therefore, the total inclusiveness loss for MyoPS-Net on labeled data can be expressed as,
\begin{equation}
    \mathcal{L}_{inc}^{L} = \mathcal{L}_{inc,S}^{L} + \mathcal{L}_{inc,E}^{L}.
\end{equation}

\subsubsection{Pathology inclusiveness loss for unlabeled data}
\subsection{Overall loss}
\label{3.4}
\begin{table*}[!t]
    \setlength{\abovecaptionskip}{0px}
    \setlength{\belowcaptionskip}{2px}
    \renewcommand{\arraystretch}{1.2}
    \caption{\label{practical}The detailed settings of four scenarios for practical usage of the proposed MyoPS-Net. The last two columns $D_\text{scar}$ and $D_\text{edema}$ represent the sets of scar decoders and edema decoders, respectively.}
    \centering
    \resizebox{0.99\linewidth}{!}{
    \begin{tabular}{p{2.5cm}|p{1.5cm}<{\centering}p{2cm}<{\centering}p{2.5cm}<{\centering}|p{6.5cm}|p{3cm}}
         \toprule
         \multirow{2}*{Model} & \multicolumn{3}{c|}{Training/ Test Data} & \multirow{2}*{$D_\text{scar}$} & \multirow{2}*{$D_\text{edema}$}\\
         \cline{2-4}
    	  ~ & All CMRs & \{C0,T2,LGE\} & \{C0,T2,mappings\} & ~ & ~\\
        \midrule
        MyoPS-Net & $\checkmark$ & $\times$ & $\times$ & \{LGE scar decoder, mappings scar decoder\} & \{T2 edema decoder\}\\
        MyoPS-Net-L & $\times$ & $\checkmark$ & $\times$ & \{LGE scar decoder\} & \{T2 edema decoder\}\\
        MyoPS-Net-M & $\times$ & $\times$ & $\checkmark$ & \{Mappings scar decoder\} & \{T2 edema decoder\}\\
        MyoPS-Net-mix & $\checkmark$ & $\checkmark$ & $\checkmark$ & \{LGE scar decoder, mappings scar decoder\} & \{T2 edema decoder\}\\
        \bottomrule
    \end{tabular}}
\end{table*}
The ground truth of myocardial scars and edema may be missing due to time-consuming and laborious manual delineation. Therefore, the inclusiveness loss can be modified to operate between the predictions for these unlabeled data ($U$),
\begin{align}
    \label{eq:incloss:U}
    \mathcal{L}_{inc,S}^{U} & = - \sum_{s \in D_\text{scar}} \sum_{e \in D_\text{edema}}  \frac{\sum_{i \in \Omega} (1-\psi_{e}^\text{edema})\log (1-\psi_{s}^\text{scar})}{\sum_{i \in \Omega} (1-\psi_{e}^\text{edema})},\\
    \mathcal{L}_{inc,E}^{U} & = - \sum_{e \in D_\text{edema}} \sum_{s \in D_\text{scar}}  \frac{\sum_{i \in \Omega} \psi_{s}^\text{scar}\log \psi_{e}^\text{edema}}{\sum_{i \in \Omega} \psi_{s}^\text{scar}},
\end{align}
where $\mathcal{L}_{inc,S}^{U}$ and $\mathcal{L}_{inc,E}^{U}$ are respectively for scar and edema decoders and then the inclusiveness loss for unlabeled data is formulated as follows,
\begin{equation}
    \mathcal{L}_{inc}^{U} = \mathcal{L}_{inc,S}^{U} + \mathcal{L}_{inc,E}^{U}.
\end{equation}

With this loss, the pathology networks can effectively perceive the spatial relationship between scars and edema for both labeled and unlabeled data.

With the above three modules, our proposed framework can then be flexibly applied to both labeled and unlabeled data.

\subsubsection{Overall loss for labeled data}
The overall loss function for labeled data is defined as,
\begin{align}
\label{all:L}
    \mathcal{L}^{L} = \mathcal{L}_{seg} + \lambda_{con} \mathcal{L}_{con} + \lambda_{inc} \mathcal{L}_{inc}^{L},
\end{align}
where $\lambda_{con}$ and $\lambda_{inc}$ are two balancing parameters. 
Besides, $\mathcal{L}_{seg}$ is a segmentation loss,
\begin{align}
\label{seg}
    \mathcal{L}_{seg} = \mathcal{L}_{seg}^{\text{MPC}} + \sum_{s \in D_\text{scar}}\lambda_{s} \mathcal{L}_{seg}^{s} + \sum_{e \in D_\text{edema}}\lambda_{e} \mathcal{L}_{seg}^{e},
\end{align}
where $\lambda_{s}$ and $\lambda_{e}$ are the hyper-parameters. 
For each term in $\mathcal{L}_{seg}$, we employ a combination of the Dice loss and the weighted cross entropy (WCE) loss \citep{sudre2017generalised}. 

\subsubsection{Overall loss for unlabeled data}
Due to a lack of ground truth, the segmentation loss is dismissed for unlabeled data. Therefore, the overall loss function for unlabeled data has the following formulation,
\begin{align}
\label{all:U}
    \mathcal{L}^{U} = \lambda_{con} \mathcal{L}_{con} + \lambda_{inc} \mathcal{L}_{inc}^{U},
\end{align}
where $\lambda_{con}$ and $\lambda_{inc}$ are also the balancing parameters.

Note that this proposed framework mainly focuses on the fully supervised situation which only has labeled data, while a preliminary study for the semi-supervised situation with both labeled and unlabeled data will be shown in Section~\ref{4.5}.

\subsection{Application of MyoPS-Net in practical clinics}
\label{3.5}
Ideally, all the five CMR sequences are available for accurate MyoPS, but in practical scenarios some sequences may be missing. 
Specifically, mapping CMR images are often excluded from a standard protocol, due to the limited scanning time. 
Also, due to the injection of contrast agents, many subjects do not take LGE CMR, for example for healthy individuals undergoing a physical examination or for patients with a kidney condition not suitable for gadolinium contrast agents.
Therefore, there are four scenarios for practical usage of the proposed MyoPS-Net, as follows.
\begin{enumerate}
    \item[(1)] \textbf{MyoPS-Net}: A model trained on full five-sequence CMR and tested solely on five sequence CMR images for MyoPS, which is also referred to as MyoPS-Net-F.
    Under this setting, the decoder set $D_\text{scar}$ contains two scar decoders respectively for LGE and mapping CMR, and the decoder set $D_\text{edema}$ contains an edema decoder for T2 CMR.
    \item[(2)] \textbf{MyoPS-Net-L}: A model trained on three-sequence CMR, \textit{i.e.,} bSSFP C0, LGE and T2 CMR, and tested solely on data of such three-sequence CMR images. 
    The decoder set $D_\text{scar}$ contains only one scar decoder for LGE CMR, and the decoder set $D_\text{edema}$ has the same setting as the first scenario.
    \item[(3)] \textbf{MyoPS-Net-M}: A model trained on four-sequence CMR, \textit{i.e.,} bSSFP C0, T2, T1 mapping and T2* mapping CMR, and tested solely on data of such four-sequence CMR images. 
    Similarly, the decoder set $D_\text{scar}$ contains a scar decoder for mapping CMR, and the decoder set $D_\text{edema}$ remains unchanged.
    \item[(4)] \textbf{MyoPS-Net-mix}: A unified model trained on a dataset consisting of all the above three combinations of multi-sequence CMR, and tested on data from any of the above three scenarios. 
    The decoder sets $D_\text{scar}$ and $D_\text{edema}$ should be modified to the correct forms corresponding to the above three situations.
\end{enumerate}
The detailed network architectures of MyoPS-Net-L, MyoPS-Net-M and MyoPS-Net-mix are shown in the Supplementary Material.
Note that when more CMR sequences are missing, such combinations can be meaningless and misleading for MyoPS in clinical practice. For example, we have the combination of bSSFP C0 and LGE CMR images for segmentation. From clinical view, this combination does not provide pathological information to edema, which may reduce the clinical meaning or pathological sense of the model. Therefore, we only considered the above four scenarios of missing either LGE CMR or mapping CMR shown in Table~\ref{practical}.

\section{Experiments and results} %%%%% 4 Experiments and results %%%%%
\label{section4}
We first illustrated the effectiveness of pathological information from each CMR sequence for MyoPS using two widely used segmentation networks (Section~\ref{4.2}). 
Then, we performed four studies to evaluate the effective setting of pathology decoders in the proposed cross-modal feature fusion architecture (Section~\ref{4.3}),
the ablated performance of each module in MyoPS-Net (Section~\ref{4.4}), the application of MyoPS-Net in the semi-supervised situation (Section~\ref{4.5}) and the applicability of MyoPS-Net to practical clinical scenarios with complex combinations of CMR sequences (Section~\ref{4.6}). 
Finally, we validated the proposed method and further applied it to a public dataset with state-of-the-art results for comparisons (Section~\ref{4.7}).

\subsection{Data and experiment setup}
\label{4.1}
\begin{table*}[!t]
    \setlength{\abovecaptionskip}{0px}
    \setlength{\belowcaptionskip}{2px}
    \renewcommand{\arraystretch}{1.2}
    \caption{\label{dataset}The acquisition parameters of the private MS-CMR dataset.}
    \centering
    \resizebox{0.75\linewidth}{!}{
    \begin{tabular}{p{2.5cm}|p{4.5cm}p{4.5cm}p{3.5cm}}
        \toprule
         Sequence & Imaging type & Slice spacing (thickness + gap) & In-plane resolution\\
        \midrule
        C0 & Cine sequence (ED phase) & $8$ mm + $2$ mm & $1.77 \times 1.77$ mm\\
        LGE & T1-weighted & $8$ mm + $2$ mm & $1.33 \times 1.33$ mm\\
        T2 & T2-weighted, black blood & $8$ mm + $2$ mm & $1.33 \times 1.33$ mm\\
        T1 mapping & T1-weighted & $8$ mm + $2$ mm & $1.41 \times 1.41$ mm\\
        T2* mapping & T2-weighted & $8$ mm + $2$ mm & $2.08 \times 2.08$ mm\\
        \bottomrule
    \end{tabular}}
\end{table*}

\begin{table}[!t]
    \setlength{\abovecaptionskip}{0px}
    \setlength{\belowcaptionskip}{2px}
    \renewcommand{\arraystretch}{1.2}
    \caption{Dice scores of myocardial pathology segmentation on each CMR sequence (LGE, T2, T1 mapping and T2* mapping CMR). The three bars in bold denote which sequence is utilized in the experiments.}\label{monoseqtabel}
    \centering
    \resizebox{0.85\linewidth}{!}{
        \begin{tabular}{p{2cm}|p{2.7cm}|p{2.7cm}}
             \toprule
             {Method/ Dice} & {Scar} & {Edema}\\
             \midrule
             \multicolumn{3}{l}{\textbf{Sequence: LGE CMR}}\\
             \midrule
             {U-Net} & {0.558 $\pm$ 0.128} & {0.563 $\pm$ 0.120}\\
             {UNet++} & {0.541 $\pm$ 0.151} & {0.582 $\pm$ 0.144}\\
             \midrule
             \multicolumn{3}{l}{\textbf{Sequence: T2 CMR}}\\
             \midrule
             {U-Net} & {0.401 $\pm$ 0.228} & {0.618 $\pm$ 0.207}\\
             {UNet++} & {0.362 $\pm$ 0.165} & {0.599 $\pm$ 0.158}\\
             \midrule
             \multicolumn{3}{l}{\textbf{Sequence: mapping (T1 mapping and T2* mapping) CMR}}\\
             \midrule
             {U-Net} & {0.352 $\pm$ 0.178} & {0.486 $\pm$ 0.167}\\
             {UNet++} & {0.353 $\pm$ 0.187} & {0.455 $\pm$ 0.180}\\
             \bottomrule
        \end{tabular}
    }
\end{table}

The private dataset employed in the experiments consists of 50 paired MS-CMR images from five sequences (bSSFP, LGE, T2, T1 mapping and T2* mapping CMR). Table~\ref{dataset} shows the detailed acquisition parameters for each sequence. 
These five sequences were aligned to a common space via the MvMM method \citep{zhuang2016multivariate,zhuang2019multivariate}. 
The in-plane resolution was resampled to 1 $\times$ 1 mm. 
The average number of slices each image is $3.64 \pm 1.09$, with the in-plane size ranging from $275 \times 339$ to $408 \times 408$.
The ground truth segmentation of scars and edema was constructed from manual annotations by one cardiologist. To calculate the  intra-observer variations, this cardiologist annotated ten randomly selected cases from this dataset once again. The intra-observer variations in terms of Dice were 0.746 $\pm$ 0.062 and 0.812 $\pm$ 0.042 for myocardial scar and edema segmentation, respectively. Besides, we invited another cardiologist to delineate the same ten cases for the inter-observer variations, which were 0.704 $\pm$ 0.092 and 0.779 $\pm$ 0.045 for myocardial scar and edema, respectively.

We randomly split the data into training ($25$ subjects), validation ($5$ subjects) and test ($20$ subjects) sets. 
All the paired images were divided into 2D slices as the network inputs, which were cropped into a unified size of $192 \times 192$ with a Z-score normalization. 
Data augmentation strategies, such as random rotation, random flip and random crop, were utilized to enrich the diversity of training images. 

We employ five evaluation metrics in the experiments, \textit{i.e.,} Dice score, Hausdorff distance (HD), Accuracy (ACC), Sensitivity (SEN) and Specificity (SPE).
\begin{align}
    \text{Dice}(S_{pre},S_{gd}) = \frac{2 \left|S_{pre} \cap S_{gd}\right|}{\left|S_{pre}\right|+\left|S_{gd}\right|},
\end{align}
where $S_{pre}$ is the prediction result and $S_{gd}$ is the ground truth.
\begin{align}
    \text{ACC} &= \frac{\text{TP}+\text{TN}}{\text{TP}+\text{FP}+\text{FN}+\text{TN}},\\
    \text{SEN} &= \frac{\text{TP}}{\text{TP}+\text{FN}},\\
    \text{SPE} &= \frac{\text{TN}}{\text{TN}+\text{FP}},
\end{align}
where $\text{TP}$ and $\text{FP}$ are the number of pixels of the true and false positives, which represent the myocardial pathologies.
Besides, $\text{TN}$ and $\text{FN}$ refer to the true and false negatives, respectively. 
ACC evaluates the overall pixel-wise accuracy of the segmentation results. 
SEN and SPE measure the ability of models to identify positive and negative pixels, respectively.

The proposed MyoPS-Net was implemented in PyTorch. 
We employed the Adam optimizer to update the network parameters (weight decay = $0.0005$). 
The initial learning rate was set to $0.0001$ and was adjusted via a cosine annealing strategy with a cycle of $20$ epochs.
The balancing parameters were set as follows, $\lambda_{s} = 2 \ (\forall s \in D_\text{scar})$, $\lambda_{e} = 2 \ (\forall e \in D_\text{edema})$ and $\lambda_{con} = 1$, $\lambda_{inc} = 1$. 
Our experiments were conducted with a computer with 2.40 GHz Intel(R) Xeon(R) Silver 4214R CPU and a 24G NVIDIA TITAN RTX GPU for 200 epochs. Moreover, the inference of our network required approximately $2.8$ seconds to predict the segmentation of one test case. 

\subsection{Results on single CMR sequence} 
\label{4.2}
\begin{figure}[!t]
    \centering
    \includegraphics[width=0.99\linewidth]{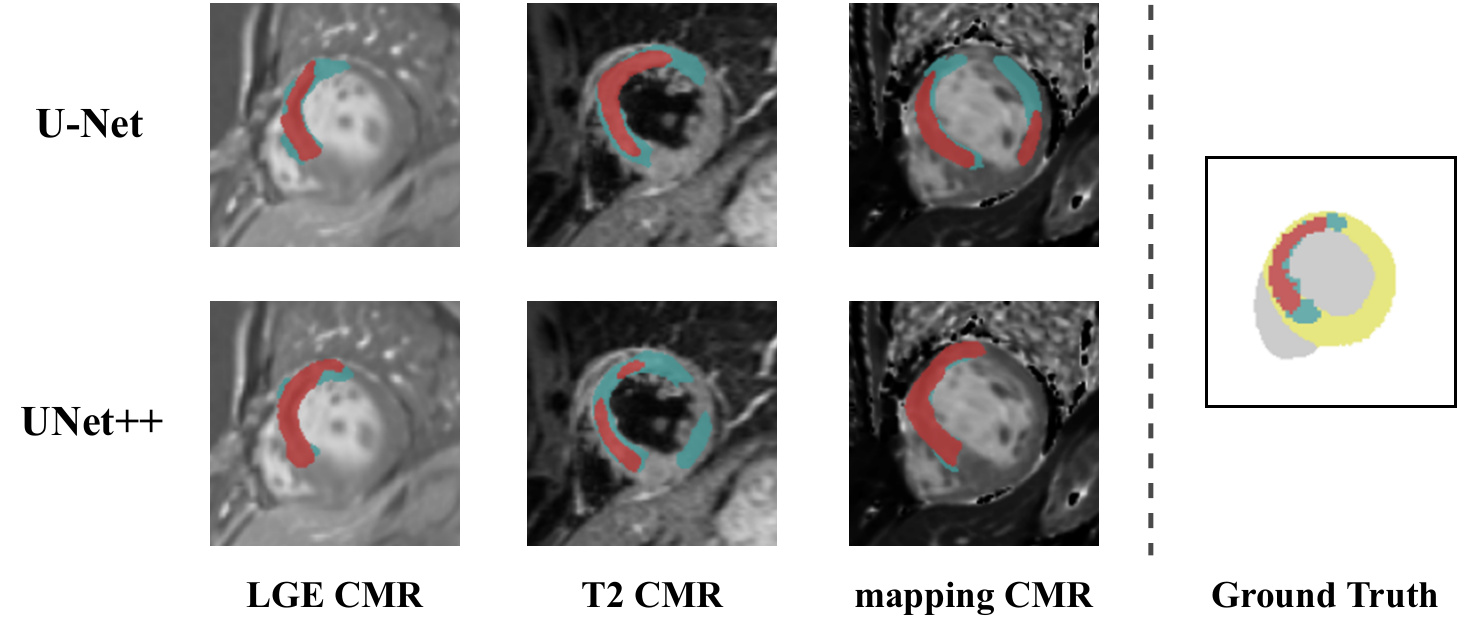}\\[-2ex]
    \caption{Visualization of MyoPS results on each CMR sequence (LGE, T2, T1 mapping and T2* mapping CMR). Specially, the segmentation results of mapping CMR are superimposed on T1 mapping CMR images. Note that scars and edema are annotated in dark red and dark green, respectively.}
    \label{monoseqfig}
\end{figure}

\begin{figure*}[t]
\centering
\includegraphics[width=0.85\textwidth]{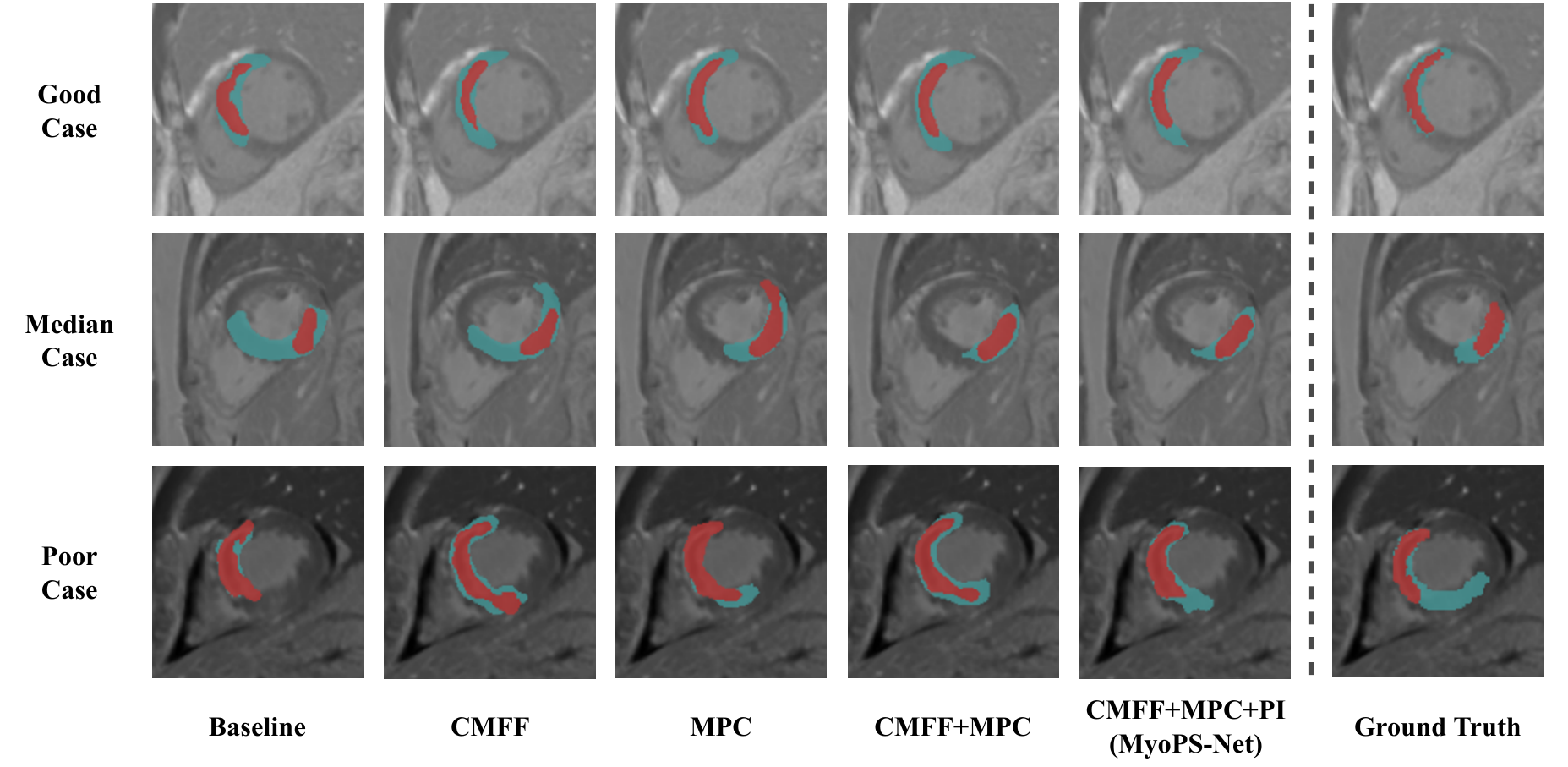}\\[-2ex]
\caption{Visualization of MyoPS results superimposed on LGE CMR from the ablation study. We selected three cases for visualization on the ground of $1/4$, $2/4$, $3/4$ of the average Dice scores, namely a good case, a median case and a poor case. Myocardial scars and edema are delineated in dark red and dark green, respectively.}
\label{ablationfig}
\end{figure*}

\begin{table*}[!t]
    \setlength{\abovecaptionskip}{0px}
    \setlength{\belowcaptionskip}{2px}
    \renewcommand{\arraystretch}{1.2}
    \caption{Quantitative evaluation results of the cross-modal feature fusion architecture under different settings of pathology decoders which are detailed in Section~\ref{4.3}. \textbf{Bold} indicates the best performance. Asterisk ($^*$) denotes the statistically significant differences given by a Wilcoxon signed-rank test with $p \leq 0.05$ between our proposed model \#5 and each comparison model.}\label{Paramter}
    \centering
    \resizebox{0.99\linewidth}{!}{
        \begin{tabular}{p{6cm}|p{2.2cm}p{2cm}p{2.2cm}p{2.2cm}p{2.2cm}|p{2.2cm}p{2cm}p{2.2cm}p{2.2cm}p{2.2cm}}
             \toprule
             \multirow{2}*{Model Architecture} & \multicolumn{5}{c|}{Scar} & \multicolumn{5}{c}{Edema}\\
             \cline{2-11}
        	  ~ & Dice & HD (mm) & ACC & SEN & SPE & Dice & HD (mm) & ACC & SEN & SPE \\
             \midrule
             {\#1. Two decoders for all sequences} & {0.607 $\pm$ 0.139 $^*$} & {12.6 $\pm$ 7.98} & {0.873 $\pm$ 0.060} & {0.583 $\pm$ 0.170 $^*$} & {0.939 $\pm$ 0.053} & {0.729 $\pm$ 0.112} & {18.8 $\pm$ 9.67} & {0.823 $\pm$ 0.071} & {0.769 $\pm$ 0.182} & {\textbf{0.858 $\pm$ 0.069}}\\
             {\#2. Six ($2\times3$) decoders for 3 sequences} & {0.606 $\pm$ 0.122 $^*$} & {12.3 $\pm$ 8.11} & {0.872 $\pm$ 0.060 $^*$} & {0.573 $\pm$ 0.167 $^*$} & {\textbf{0.946 $\pm$ 0.045}} & {0.723 $\pm$ 0.075} & {22.1 $\pm$ 12.4} & {0.807 $\pm$ 0.067 $^*$} & {0.793 $\pm$ 0.163} & {0.822 $\pm$ 0.082 $^*$}\\
             {\#3. Four decoders (two for mappings)} & {0.631 $\pm$ 0.147} & {13.1 $\pm$ 7.62} & {0.873 $\pm$ 0.068} & {\textbf{0.631 $\pm$ 0.208}} & {0.934 $\pm$ 0.057 $^*$} & {0.708 $\pm$ 0.112 $^*$} & {20.0 $\pm$ 11.4} & {0.794 $\pm$ 0.083 $^*$} & {0.802 $\pm$ 0.153} & {0.801 $\pm$ 0.108 $^*$}\\
             {\#4. Three decoders (edema for mappings)} & {0.626 $\pm$ 0.126} & {12.4 $\pm$ 8.24} & {0.876 $\pm$ 0.057 $^*$} & {0.618 $\pm$ 0.164} & {0.939 $\pm$ 0.039} & {0.721 $\pm$ 0.111} & {20.7 $\pm$ 12.1} & {0.805 $\pm$ 0.079} & {\textbf{0.813 $\pm$ 0.158}} & {0.807 $\pm$ 0.097 $^*$}\\
             \midrule
             {\#5. Three decoders (scar for mappings)} & {\textbf{0.656 $\pm$ 0.113}} & {\textbf{11.4 $\pm$ 9.45}} & {\textbf{0.886 $\pm$ 0.049}} & {0.626 $\pm$ 0.135} & {\textbf{0.946 $\pm$ 0.045}} & {\textbf{0.741 $\pm$ 0.085}} & {\textbf{18.6 $\pm$ 10.6}} & {\textbf{0.829 $\pm$ 0.073}} & {0.775 $\pm$ 0.160} & {\textbf{0.858 $\pm$ 0.073}}\\
             \bottomrule
        \end{tabular}
    }
\end{table*}

To illustrate the relative importance of each CMR sequence (LGE, T2, T1 mapping and T2* mapping CMR) for MyoPS, we adopted U-Net \citep{unet} and UNet++ \citep{unet++} to conduct these experiments. 
Table~\ref{monoseqtabel} shows the quantitative results in terms of Dice, and Fig.~\ref{monoseqfig} visualizes the segmentation results of one randomly selected case. These models employed 25 cases from each CMR sequence for training and were tested on 20 cases from the same CMR sequence.

As for LGE CMR, both networks achieved evidently better Dice scores on scar segmentation than the other two sequences. 
Besides, the Dice scores of edema on LGE CMR were also good, thanks to the accurate scar segmentation which accounts for a major proportion of the edema areas.
For T2 CMR, its edema segmentation was reliable by both two models, but the scar segmentation had a great drop in Dice scores. 
This is reasonable as T2 CMR usually provides edema information while LGE CMR mainly offers scar-related information. 
For mapping CMR, the segmentation of either two types of pathologies was not accurate enough, thus more information from other sequences would be needed for reliable MyoPS.

\subsection{Ablation study on the model architecture}
\label{4.3}
\begin{figure*}[t]
\centering
      \centering
    \subfigure[]{
    \label{result:a}
    \includegraphics[width=0.45\textwidth]{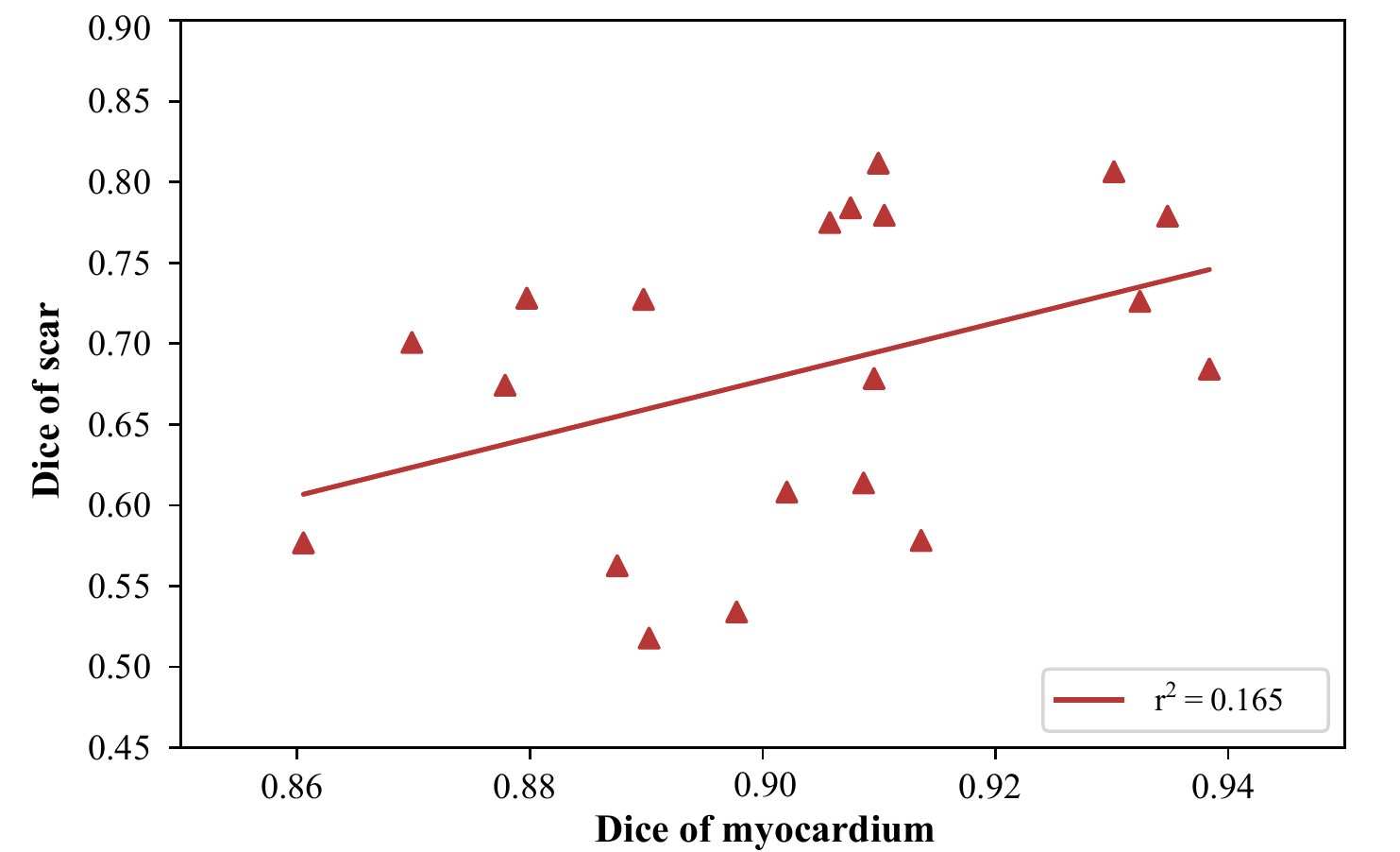}}
     \subfigure[]{
       \label{result:b}
        \includegraphics[width=0.45\textwidth]{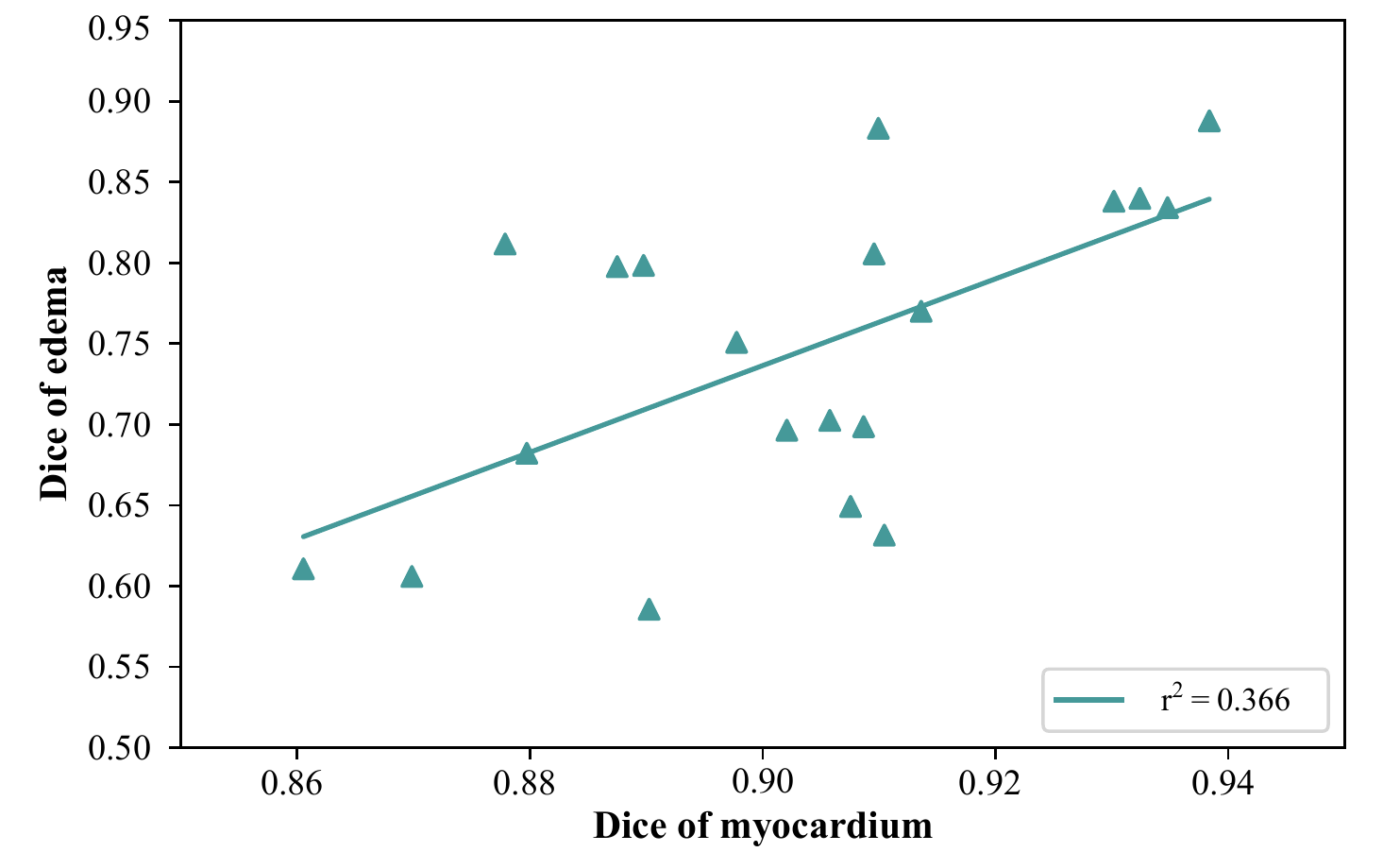}}
        \caption{The scatter point plots and correlation between the Dice scores of the myocardium and the Dice scores of the pathologies (scars and edema in (a) and (b), respectively).}
\label{corr}
\end{figure*}

\begin{table*}[!t]
  \setlength{\abovecaptionskip}{0px}
  \setlength{\belowcaptionskip}{2px}
  \renewcommand{\arraystretch}{1.2}
  \caption{Quantitative evaluation results of ablation study on the proposed modules. CMFF: cross-modal feature fusion module. MPC: myocardium prior and consistency module. PI: prior for pathology inclusiveness. \textbf{Bold} indicates the best performance. Asterisk ($^*$) denotes the statistically significant improvement ($p \leq 0.05$) of each model compared with the baseline.}\label{ablationtable}
  \begin{center}
  \resizebox{0.99\textwidth}{!}{
    \begin{tabular}{p{5cm}|p{2.2cm}p{2cm}p{2.2cm}p{2.2cm}p{2.2cm}|p{2.2cm}p{2cm}p{2.2cm}p{2.2cm}p{2.2cm}}
     \toprule
     \multirow{2}*{Models} & \multicolumn{5}{c|}{Scar} & \multicolumn{5}{c}{Edema}\\
     \cline{2-11}
	  ~ & Dice & HD (mm) & ACC & SEN & SPE & Dice & HD (mm) & ACC & SEN & SPE \\
     \midrule
     {Baseline} & {0.560 $\pm$ 0.133} & {18.7 $\pm$ 14.4} & {0.849 $\pm$ 0.055} & {0.595 $\pm$ 0.182} & {0.907 $\pm$ 0.058} & {0.668 $\pm$ 0.152} & {21.8 $\pm$ 14.0} & {0.789 $\pm$ 0.082} & {0.714 $\pm$ 0.202} & {0.824 $\pm$ 0.078}\\
     {Baseline+CMFF} & {0.614 $\pm$ 0.137 $^*$} & {14.3 $\pm$ 8.66} & {0.868 $\pm$ 0.056 $^*$} & {0.621 $\pm$ 0.169} & {0.925 $\pm$ 0.063 $^*$} & {0.721 $\pm$ 0.099 $^*$} & {19.0 $\pm$ 11.2} & {0.818 $\pm$ 0.075 $^*$} & {0.744 $\pm$ 0.177} & {0.861 $\pm$ 0.079 $^*$}\\
     {Baseline+MPC} & {0.625 $\pm$ 0.132 $^*$} & {20.8 $\pm$ 14.9} & {0.843 $\pm$ 0.059} & {\textbf{0.768 $\pm$ 0.112} $^*$} & {0.865 $\pm$ 0.062} & {0.717 $\pm$ 0.118} & {25.2 $\pm$ 12.6} & {0.800 $\pm$ 0.060} & {\textbf{0.823 $\pm$ 0.135} $^*$} & {0.795 $\pm$ 0.075}\\
     {Baseline+CMFF+MPC} & {0.627 $\pm$ 0.119 $^*$} & {13.7 $\pm$ 9.80} & {0.870 $\pm$ 0.058 $^*$} & {0.634 $\pm$ 0.152} & {0.927 $\pm$ 0.047 $^*$} & {0.733 $\pm$ 0.111 $^*$} & {19.4 $\pm$ 10.5} & {0.825 $\pm$ 0.083 $^*$} & {0.742 $\pm$ 0.175 $^*$} & {\textbf{0.879 $\pm$ 0.065} $^*$}\\
     \midrule
     {Baseline+CMFF+MPC+PI (Ours)} & {\textbf{0.656 $\pm$ 0.113} $^*$} & {\textbf{11.4 $\pm$ 9.45}} & {\textbf{0.886 $\pm$ 0.049} $^*$} & {0.626 $\pm$ 0.135} & {\textbf{0.946 $\pm$ 0.045} $^*$} & {\textbf{0.741 $\pm$ 0.085} $^*$} & {\textbf{18.6 $\pm$ 10.6}} & {\textbf{0.829 $\pm$ 0.073} $^*$} & {0.775 $\pm$ 0.160} & {0.858 $\pm$ 0.073 $^*$}\\
     \bottomrule
  \end{tabular}}
  \end{center}
\end{table*}

\begin{table*}[!t]
    \setlength{\abovecaptionskip}{0px}
    \setlength{\belowcaptionskip}{2px}
    \renewcommand{\arraystretch}{1.2}
    \caption{Quantitative evaluation results of the application of MyoPS-Net in the semi-supervised situation. The columns \textit{Number of data} indicate the specific number of labeled data and unlabeled data in each setting. Asterisk ($^*$) denotes the statistically significant differences ($p \leq 0.05$) between each semi-supervised model and the corresponding fully supervised model.}\label{semi-supervised}
    \centering
    \resizebox{0.99\linewidth}{!}{
        \begin{tabular}{p{1.5cm}<{\centering}|p{1.5cm}<{\centering}|p{2.2cm}p{2cm}p{2.2cm}p{2.2cm}p{2.2cm}|p{2.2cm}p{2cm}p{2.2cm}p{2.2cm}p{2.2cm}}
             \toprule
             \multicolumn{2}{c|}{Number of data} & \multicolumn{5}{c|}{Scar} & \multicolumn{5}{c}{Edema}\\
             \cline{1-12}
        	 Labeled & Unlabeled & Dice & HD (mm) & ACC & SEN & SPE & Dice & HD (mm) & ACC & SEN & SPE \\
             \midrule
             {25} & {0} & {0.656 $\pm$ 0.113} & {11.4 $\pm$ 9.45} & {0.886 $\pm$ 0.049} & {0.626 $\pm$ 0.135} & {0.946 $\pm$ 0.045} & {0.741 $\pm$ 0.085} & {18.6 $\pm$ 10.6} & {0.829 $\pm$ 0.073} & {0.775 $\pm$ 0.160} & {0.858 $\pm$ 0.073}\\
             \midrule
             {20} & {5} & {0.645 $\pm$ 0.115} & {16.5 $\pm$ 10.5} & {0.873 $\pm$ 0.054 $^*$} & {0.725 $\pm$ 0.134} & {0.909 $\pm$ 0.061 $^*$} & {0.729 $\pm$ 0.101 $^*$} & {20.6 $\pm$ 11.3} & {0.808 $\pm$ 0.083 $^*$} & {0.812 $\pm$ 0.147} & {0.811 $\pm$ 0.115 $^*$}\\
             {20} & {0} & {0.628 $\pm$ 0.125} & {17.4 $\pm$ 12.9} & {0.848 $\pm$ 0.069} & {0.744 $\pm$ 0.166} & {0.876 $\pm$ 0.074} & {0.697 $\pm$ 0.088} & {24.6 $\pm$ 10.5} & {0.768 $\pm$ 0.074} & {0.853 $\pm$ 0.105} & {0.738 $\pm$ 0.113}\\
             \midrule
             {15} & {10} & {0.631 $\pm$ 0.115 $^*$} & {15.1 $\pm$ 10.8} & {0.862 $\pm$ 0.062} & {0.686 $\pm$ 0.163 $^*$} & {0.903 $\pm$ 0.072} & {0.694 $\pm$ 0.087} & {27.2 $\pm$ 12.0} & {0.757 $\pm$ 0.102} & {0.833 $\pm$ 0.109} & {0.717 $\pm$ 0.145}\\
             {15} & {0} & {0.584 $\pm$ 0.122} & {18.1 $\pm$ 12.4} & {0.857 $\pm$ 0.064} & {0.593 $\pm$ 0.181} & {0.921 $\pm$ 0.067} & {0.676 $\pm$ 0.106} & {24.9 $\pm$ 12.8} & {0.758 $\pm$ 0.098} & {0.788 $\pm$ 0.172} & {0.756 $\pm$ 0.132}\\
             \midrule
             {10} & {15} & {0.626 $\pm$ 0.124 $^*$} & {15.1 $\pm$ 11.4} & {0.856 $\pm$ 0.074} & {0.683 $\pm$ 0.163 $^*$} & {0.893 $\pm$ 0.092} & {0.685 $\pm$ 0.094} & {26.3 $\pm$ 12.3} & {0.759 $\pm$ 0.091} & {0.822 $\pm$ 0.131 $^*$} & {0.727 $\pm$ 0.129}\\
             {10} & {0} & {0.566 $\pm$ 0.126} & {21.6 $\pm$ 13.2} & {0.849 $\pm$ 0.060} & {0.586 $\pm$ 0.192} & {0.910 $\pm$ 0.075} & {0.669 $\pm$ 0.123} & {26.7 $\pm$ 12.3} & {0.742 $\pm$ 0.123} & {0.788 $\pm$ 0.156} & {0.722 $\pm$ 0.152}\\
             \bottomrule
        \end{tabular}
    }
\end{table*}

To evaluate the effectiveness of the model architecture, we conducted experiments on four different settings of pathology decoders in the proposed cross-modal feature fusion architecture. 
Table~\ref{Paramter} illustrates the quantitative evaluation results of these four frameworks along with the proposed MyoPS-Net. 
Model \#1 employs two decoders (a scar decoder and an edema decoder) for all CMRs to perform MyoPS. 
Specifically, the fused features from the four CMR sequences except C0 were fed into the scar decoder and the edema decoder for pathology segmentation via a pixel-wise max operation.
The detailed structure of Model \#1 is presented in the Supplementary Material.
Model \#2 indicates the basic situation of six decoders for three CMRs with a scar decoder and an edema decoder for each. 
Model \#3,  \#4 and \# 5 have the same setting on LGE and T2 CMR as the proposed MyoPS-Net, namely a scar decoder for LGE CMR and an edema decoder for T2 CMR. 
Differently, model \#3 holds two decoders (a scar decoder and an edema decoder) for mapping CMR, model \#4 has only an edema decoder, while \#5 (the proposed MyoPS-Net) is designed with a scar decoder for mapping CMR. 
As shown in Table~\ref{Paramter}, with one specific pathology decoder, model \#5 achieved the best performance.
Therefore, in the following studies, we adopted this setting of model \#5 for the proposed MyoPS-Net.

\subsection{Ablation study on the proposed modules}
\label{4.4}
\begin{table*}[!t]
    \setlength{\abovecaptionskip}{0px}
    \setlength{\belowcaptionskip}{2px}
    \renewcommand{\arraystretch}{1.2}
    \caption{Quantitative evaluation results of variants of the proposed MyoPS-Net in different scenarios of practical clinics. The upper half represents the results of one model, which was trained on full data with the setting of complex combinations of different CMR sequences and was tested on three datasets representing three different scenarios. The lower half indicates the results of the three models tested on three same test datasets. Asterisk ($^*$) denotes the statistically significant improvement ($p \leq 0.05$) of models from the same test setting.}\label{missing}
    \centering
    \resizebox{0.99\linewidth}{!}{
        \begin{tabular}{p{5.5cm}|p{2.2cm}p{2cm}p{2.2cm}p{2.2cm}p{2.2cm}|p{2.2cm}p{2cm}p{2.2cm}p{2.2cm}p{2.2cm}}
             \toprule
             \multirow{2}*{Experimental setting} & \multicolumn{5}{c|}{Scar} & \multicolumn{5}{c}{Edema}\\
             \cline{2-11}
        	  ~ & Dice & HD (mm) & ACC & SEN & SPE & Dice & HD (mm) & ACC & SEN & SPE \\
        	 \midrule
        	 \multicolumn{11}{l}{\textbf{Setting: one model, \textit{i.e.,} MyoPS-Net-mix, trained on full data with complex combinations}}\\
             \midrule
             {Tested on all CMRs} & {0.628 $\pm$ 0.138 $^*$} & {15.9 $\pm$ 9.94} & {0.872 $\pm$ 0.050 $^*$} & {0.655 $\pm$ 0.186 $^*$} & {0.920 $\pm$ 0.065} & {0.711 $\pm$ 0.102} & {21.0 $\pm$ 12.5} & {0.798 $\pm$ 0.064} & {0.804 $\pm$ 0.139 $^*$} & {0.797 $\pm$ 0.084}\\
             {Tested on \{C0,T2,LGE\}} & {0.586 $\pm$ 0.180 $^*$} & {14.9 $\pm$ 11.0} & {0.878 $\pm$ 0.049 $^*$} & {0.545 $\pm$ 0.195} & {0.948 $\pm$ 0.051 $^*$} & {0.674 $\pm$ 0.173} & {23.0 $\pm$ 15.0} & {0.798 $\pm$ 0.072} & {0.704 $\pm$ 0.218} & {0.833 $\pm$ 0.084}\\
             {Tested on \{C0,T2,mappings\}} & {0.427 $\pm$ 0.175 $^*$} & {25.6 $\pm$ 15.9} & {0.809 $\pm$ 0.066} & {0.471 $\pm$ 0.258 $^*$} & {0.886 $\pm$ 0.075} & {0.654 $\pm$ 0.148 $^*$} & {22.2 $\pm$ 14.4} & {0.760 $\pm$ 0.095} & {0.731 $\pm$ 0.190 $^*$} & {0.778 $\pm$ 0.111}\\
             \midrule
             \multicolumn{11}{l}{\textbf{Setting: three separated models trained on three datasets, each with one-third of training data}}\\
             \midrule
             {MyoPS-Net-F on all CMRs} & {0.585 $\pm$ 0.166} & {14.6 $\pm$ 9.53} & {0.844 $\pm$ 0.072} & {0.561 $\pm$ 0.226} & {0.921 $\pm$ 0.061} & {0.702 $\pm$ 0.110} & {18.0 $\pm$ 10.8} & {0.813 $\pm$ 0.091} & {0.676 $\pm$ 0.198} & {0.900 $\pm$ 0.065}\\
             {MyoPS-Net-L on \{C0,T2,LGE\}} & {0.561 $\pm$ 0.132} & {15.4 $\pm$ 11.1} & {0.854 $\pm$ 0.068} & {0.538 $\pm$ 0.177} & {0.932 $\pm$ 0.051} & {0.682 $\pm$ 0.112} & {18.7 $\pm$ 9.71} & {0.792 $\pm$ 0.091} & {0.711 $\pm$ 0.184} & {0.838 $\pm$ 0.092}\\
             {MyoPS-Net-M on \{C0,T2,mappings\}} & {0.392 $\pm$ 0.153} & {29.4 $\pm$ 14.0} & {0.792 $\pm$ 0.064} & {0.405 $\pm$ 0.208} & {0.881 $\pm$ 0.064} & {0.587 $\pm$ 0.211} & {28.3 $\pm$ 12.4} & {0.722 $\pm$ 0.124} & {0.644 $\pm$ 0.268} & {0.755 $\pm$ 0.145}\\
             \bottomrule
        \end{tabular}
    }
 \end{table*}
This study investigates the effectiveness of each separate module in the proposed MyoPS-Net. 
Table~\ref{ablationtable} presents the quantitative evaluation results of models with different modules and Fig.~\ref{ablationfig} visualizes the segmentation results of three cases based on $1/4$, $2/4$, $3/4$ of the average Dice scores, \textit{i.e.,} a good case, a median case and a poor case. 

The baseline was only with basic U-Nets, and obtained a poor performance for pathology segmentation. 
For the CMFF module, we tried two different strategies, \textit{i.e.,} max operation and mean operation, to fuse multi-sequence features. However, the average Dice scores of employing the mean operation were 0.603 $\pm$ 0.133 and 0.713 $\pm$ 0.114 for myocardial scars and edema, respectively. Based on the above results, we utilized the max operation in our model.
Then, with the CMFF module, the segmentation performance of the model was boosted, compared to the baseline with evidently and significantly better pathology segmentation.
This verifies the importance of cross-modal feature fusion architecture, which utilizes the complementary information from the multi-sequence CMR for MyoPS. 
The model only with the MPC module also achieved significantly better Dice scores than the baseline for both scar and edema segmentation.
To further demonstrate the effectiveness of the MPC module, we performed a correlation analysis between the Dice scores of the myocardium segmentation and the Dice scores of the MyoPS. As shown in Fig.~\ref{corr}, there existed approximate positive linear correlations between the myocardium and the pathologies, verifying that the MPC module assisted the framework in localizing and then segmenting the pathologies.
As shown in Fig.~\ref{ablationfig}, the segmentation results of both the model with CMFF and the model with MPC had smoother and more accurate boundaries, compared to the baseline. 
When combining the CMFF and MPC modules, the model obtained further improved accuracies. Besides, the shape of segmentation results was further optimized to be close to the ground truth.
Finally, integrating the proposed three modules, we had the proposed MyoPS-Net, which achieved the best Dice scores and HDs for MyoPS.
These segmentation results illustrate that MyoPS-Net can employ the pathology inclusiveness to further improve the MyoPS performance.

\subsection{Application in the semi-supervised situation}
\label{4.5}
 \begin{table*}[!t]
    \setlength{\abovecaptionskip}{0px}
    \setlength{\belowcaptionskip}{2px}
    \renewcommand{\arraystretch}{1.2}
    \caption{Quantitative evaluation results of the proposed MyoPS-Net, MyoPS-Net-L and MyoPS-Net-M. For reference, the result of MyoPS-Net-L on the public MICCAI2020 MyoPS Challenge dataset \citep{zhuang2016multivariate,zhuang2019multivariate} is presented at the bottom line.}\label{proposed}
    \centering
    \resizebox{0.99\linewidth}{!}{
        \begin{tabular}{p{2.5cm}|p{2cm}p{1.8cm}p{2cm}p{2cm}p{2cm}|p{2cm}p{1.8cm}p{2cm}p{2cm}p{2cm}}
             \toprule
             \multirow{2}*{Model} & \multicolumn{5}{c|}{Scar} & \multicolumn{5}{c}{Edema}\\
             \cline{2-11}
        	  ~ & Dice & HD (mm) & ACC & SEN & SPE & Dice & HD (mm) & ACC & SEN & SPE \\
             \midrule
             \multicolumn{11}{l}{\textbf{Results on private dataset}}\\
             \midrule
             {MyoPS-Net} & {0.656 $\pm$ 0.113} & {11.4 $\pm$ 9.45} & {0.886 $\pm$ 0.049} & {0.626 $\pm$ 0.135} & {0.946 $\pm$ 0.045} & {0.741 $\pm$ 0.085} & {18.6 $\pm$ 10.6} & {0.829 $\pm$ 0.073} & {0.775 $\pm$ 0.160} & {0.858 $\pm$ 0.073}\\
             {MyoPS-Net-L} & {0.622 $\pm$ 0.116} & {11.4 $\pm$ 8.17} & {0.881 $\pm$ 0.050} & {0.569 $\pm$ 0.153} & {0.952 $\pm$ 0.040} & {0.727 $\pm$ 0.102} & {21.2 $\pm$ 13.2} & {0.818 $\pm$ 0.082} & {0.763 $\pm$ 0.171} & {0.858 $\pm$ 0.084}\\
             {MyoPS-Net-M} & {0.501 $\pm$ 0.181} & {27.5 $\pm$ 14.5} & {0.810 $\pm$ 0.060} & {0.609 $\pm$ 0.224} & {0.856 $\pm$ 0.065} & {0.676 $\pm$ 0.124} & {25.1 $\pm$ 11.9} & {0.772 $\pm$ 0.075} & {0.787 $\pm$ 0.173} & {0.763 $\pm$ 0.095}\\
             \midrule
             \multicolumn{11}{l}{\textbf{Results on public dataset}}\\
             \midrule
             {MyoPS-Net-L} & {0.647 $\pm$ 0.258} & {15.5 $\pm$ 14.9} & {0.865 $\pm$ 0.089} & {0.713 $\pm$ 0.234} & {0.919 $\pm$ 0.054} & {0.722 $\pm$ 0.135} & {22.9 $\pm$ 16.2} & {0.791 $\pm$ 0.109} & {0.727 $\pm$ 0.172} & {0.827 $\pm$ 0.110}\\
             \bottomrule
        \end{tabular}
    }
\end{table*}
 
This study evaluates the application of MyoPS-Net in a semi-supervised situation. Table~\ref{semi-supervised} illustrates the quantitative evaluation results of different models employing different numbers of labeled and unlabeled data. Among these models, we experimented with three semi-supervised situations, which employ a total of 25 cases for training, with the number of labeled data decreasing from 20 to 10 in a step of 5. For comparison, we additionally conducted four experiments in a fully supervised situation, with the number of labeled data ranging from 25 to 10 in the same step.

The model with 25 labeled data achieved the best segmentation performance.
With the proportion of unlabeled data increasing, the segmentation performance of the semi-supervised models gradually dropped.
Nevertheless, the three semi-supervised models achieved comparable scar segmentation results towards the model with 25 labeled data.
The segmentation performance of edema evidently decreased when more than 10 unlabeled cases were utilized, which illustrated that edema segmentation was much more sensitive to the reduction of supervision than scar segmentation.
Besides, these semi-supervised models outperformed the corresponding fully supervised models with the additional unlabeled data.
For example, the semi-supervised model with 20 labeled data and 5 unlabeled data achieved better scar and edema segmentation than the fully supervised model only with 20 labeled data, as well as the other two situations.
These results demonstrated that the proposed MyoPS-Net can employ complementary pathological information from multi-sequence CMR images for effective semi-supervised MyoPS.

\subsection{Application to complex combinations of CMR sequences in practical clinics}
\label{4.6}
To demonstrate the effectiveness of the proposed framework on complex combinations of different CMR sequences, we conducted experiments for MyoPS-Net-mix which is presented in Section~\ref{3.5}. 
The training data of MyoPS-Net-mix consisted of three parts, \textit{i.e.,} $9$ cases with full five-sequence CMR, $8$ cases with C0, T2, LGE CMR and $8$ cases with C0, T2, mapping CMR. Therefore, a total of $25$ cases were employed to train MyoPS-Net-mix.
Besides, MyoPS-Net-mix was tested on three different sets, \textit{i.e.,} one of $20$ cases with five-sequence CMR, one of $20$ cases with C0, T2, LGE CMR and one of $20$ cases with C0, T2, mapping CMR.

For comparisons, we further implemented three separate models, \textit{i.e.,} MyoPS-Net-F, MyoPS-Net-L and MyoPS-Net-M.
The training data of MyoPS-Net-mix was employed to train the above three models respectively.
MyoPS-Net-F was trained on the set of $9$ cases with five-sequence CMR and tested on the same test set of $20$ cases as MyoPS-Net-mix.
MyoPS-Net-L was trained on the set of $8$ cases with C0, T2, LGE CMR and tested on the same $20$ cases.
Besides, MyoPS-Net-M was trained on the set of $8$ cases with C0, T2, mapping CMR and tested on the same $20$ cases.

Table~\ref{missing} illustrated the quantitative evaluation results of the proposed MyoPS-Net-mix.
The segmentation results of the proposed MyoPS-Net-mix on the scenarios of five-sequence CMR and three-sequence CMR without LGE were evidently better than the corresponding separate models. 
Besides, MyoPS-Net-mix achieved a similar segmentation performance on the scenario of four-sequence CMR without mappings compared to the separate model.
These results demonstrated that the proposed MyoPS-Net-mix can acquire complementary pathological features from different settings to prompt better segmentation performance, which is much more effective in MyoPS than separate models.

\subsection{Results of the proposed methods}
\label{4.7}
\begin{figure*}[!t]
    \centering
      \centering
    \subfigure[]{
    \label{result:a}
    \includegraphics[width=0.45\textwidth]{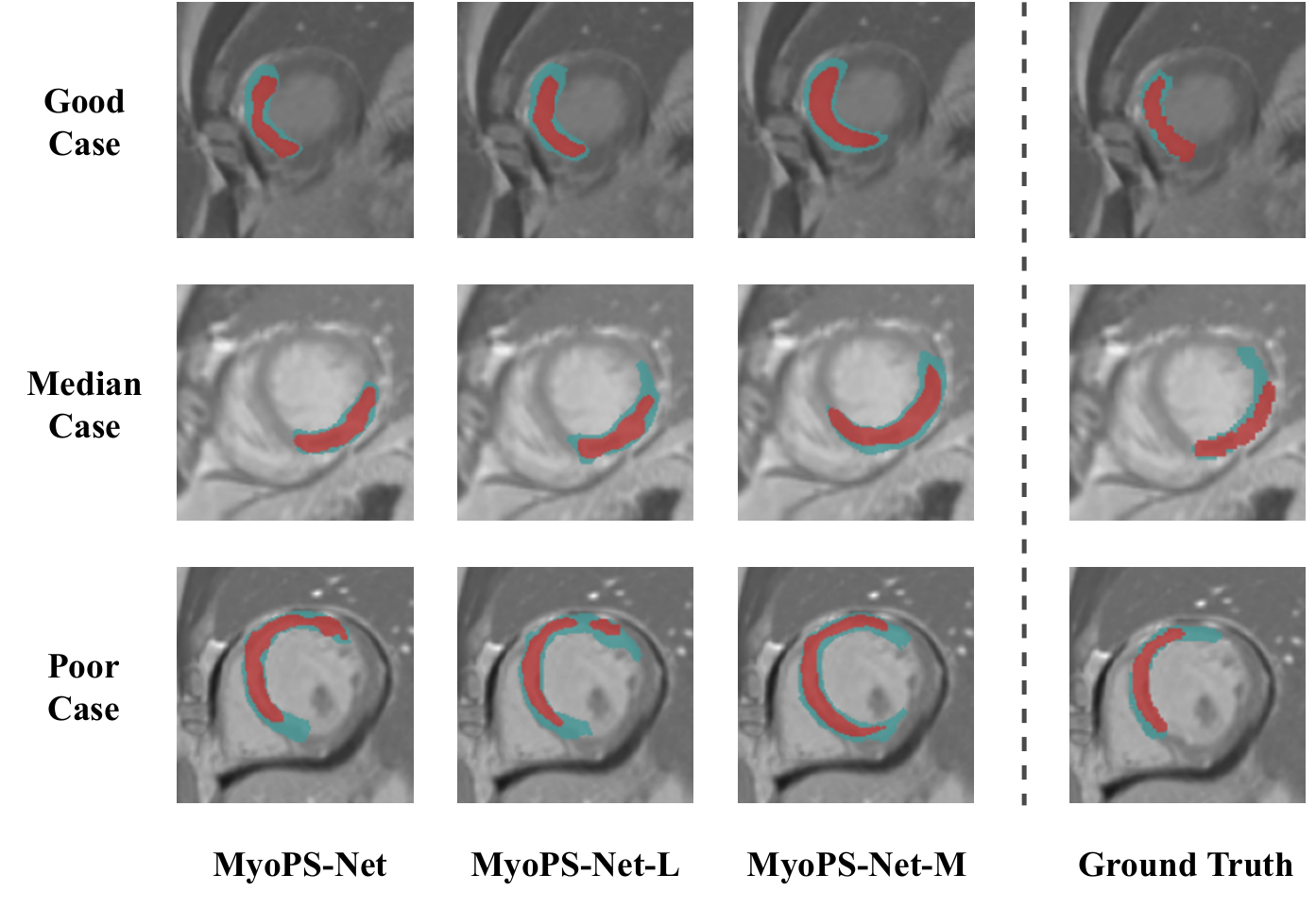}}
     \subfigure[]{
       \label{result:b}
        \includegraphics[width=0.47\textwidth]{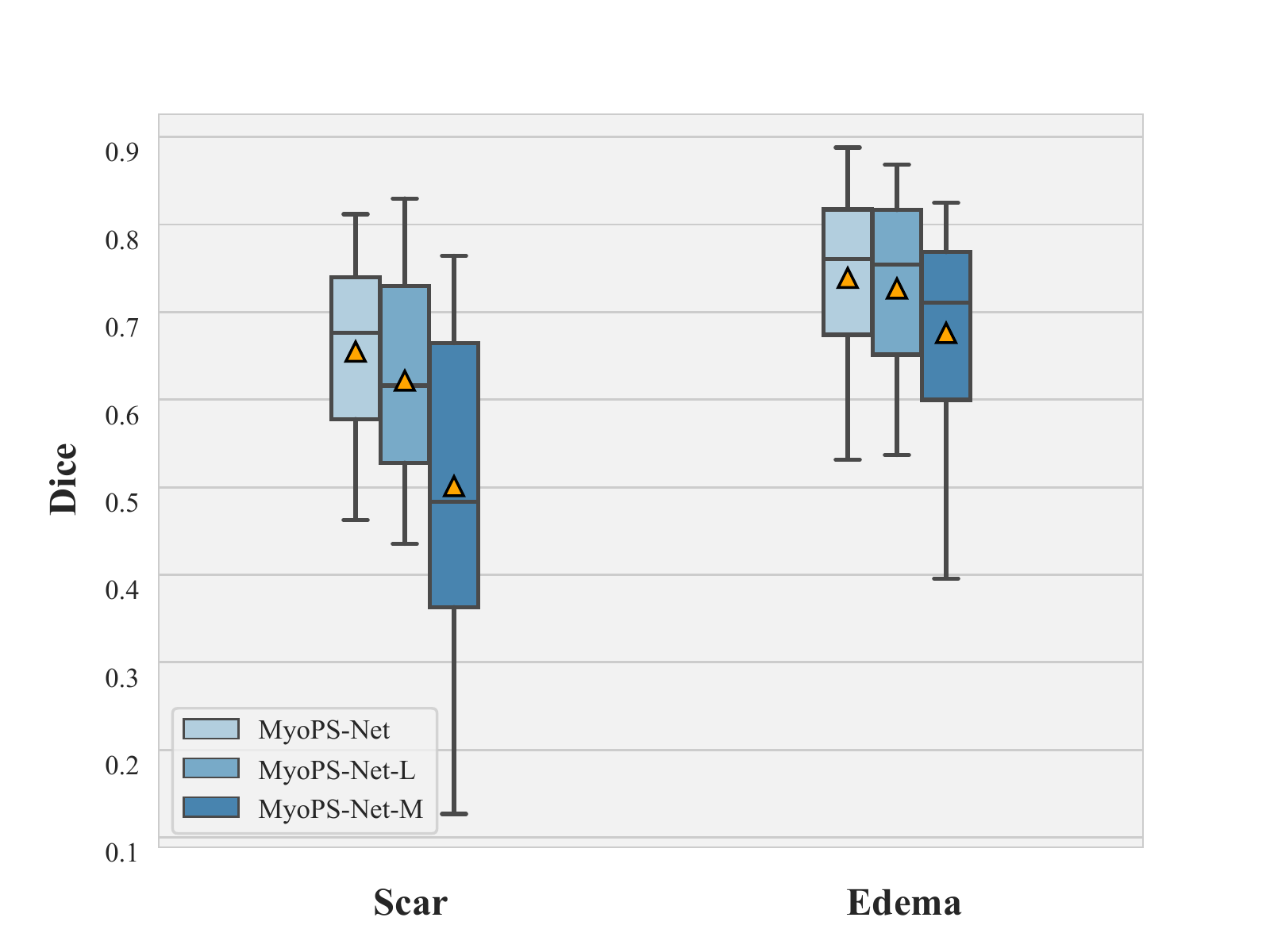}}
    \caption{(a) Visualization of MyoPS results superimposed on LGE CMR. We select three cases for visualization ranked by the average Dice scores, \textit{i.e.,} a good case ($1/4$), a median case ($2/4$) and a poor case ($3/4$). The structures in dark red denote the myocardial scar and the edema regions are represented in dark green. (b) The box plot of the Dice scores. The triangles in orange denote the average Dice scores.}
    \label{result}
\end{figure*}
To evaluate the segmentation performance of the proposed methods, we conducted three experiments respectively for the proposed MyoPS-Net, MyoPS-Net-L and MyoPS-Net-M mentioned in Section~\ref{3.5} using full $25$ training images. Table~\ref{proposed} presented the quantitative evaluation results of these three models. Fig.~\ref{result} (a) visualized the segmentation results of three typical cases in terms of $1/4$, $2/4$, $3/4$ of the average Dice scores, namely a good case, a median case and a poor case, and Fig.~\ref{result} (b) illustrated the Dice scores of three models using box plots.

All of these three models achieved promising MyoPS results on the private dataset.
For MyoPS-Net, which obtained the best Dice scores and HDs, the segmentation performance had smoother and more accurate results than the other two models, which demonstrated that MyoPS-Net could excavate instructive features from five-sequence CMR images for a better MyoPS performance.
(R2-Comment2.2) Besides, one can see that the proposed MyoPS-Net can obtain comparable results toward the intra- and inter-observer variations described in Section~\ref{4.1}.
MyoPS-Net-L achieved competitive performance without employing mapping CMR. However, the segmentation results of MyoPS-Net-L showed a decrease compared to MyoPS-Net, which illustrated that mapping CMR provides complementary pathological features for MyoPS.
Without the assistance of LGE CMR, MyoPS-Net-M achieved a relatively low but competitive performance, which revealed that this framework can effectively extract pathological information from T2 and mapping CMR to perform MyoPS. 

To further demonstrate the effectiveness of the proposed framework, an extensive experiment was conducted on the public dataset from the MICCAI2020 MyoPS challenge \citep{zhuang2016multivariate,zhuang2019multivariate}.
Table~\ref{MyoPS} presented the average Dice scores of the proposed framework compared to outstanding algorithms from the MyoPS Challenge. 
Specifically, AWSnet \citep{AWSnet} and Modified nnUNet \citep{myopschallenge-twostage-3} performed MyoPS via a two-stage strategy, while EfficientSeg \citep{EfficientSeg}, CMS-UNet \citep{CMS-UNet} and MF\&DFA-Net \citep{Dual-Path} were end-to-end learning frameworks.
Considering the scenario of this challenge, namely three-sequence CMR without mapping CMR, we employed MyoPS-Net-L to conduct an experiment with the same training and test sets as the public dataset.
Besides, ensemble learning methods have been considered to achieve superior prediction performance \citep{ensemble1}, such as bagging or boosting. We then employed the bagging strategy, one of the most popular ensemble learning methods, to reduce uncertainty and increase the generalization abilities of our model \citep{ensemble2}.
We trained multiple base models by modifying the initial weights. The final prediction was then achieved by pixel-wise majority voting on each corresponding prediction of these base models.
The experimental results proved that the proposed MyoPS-Net-L achieved comparable MyoPS performance toward the state-of-the-art results.
\begin{table}[!t]
    \setlength{\abovecaptionskip}{0px}
    \setlength{\belowcaptionskip}{2px}
  \renewcommand{\arraystretch}{1.2}
  \caption{Experimental results on the public MICCAI2020 MyoPS Challenge dataset \citep{zhuang2016multivariate,zhuang2019multivariate}. Note that the results of the compared methods were cited directly from the original publications. \textbf{Bold} indicates the best performance. The superscript ($^{**}$) indicates that the original method adopted ensemble learning in the experiments.}
  \label{MyoPS}
  \begin{center}
  \resizebox{0.99\linewidth}{!}{
    \begin{tabular}{p{6.5cm}|p{2cm}|p{2cm}|p{1cm}}
     \toprule
     Method/ Dice & \multicolumn{1}{l|}{Scar} & Edema & Avg\\ 
     \midrule
    {AWSnet $^{**}$ \citep{AWSnet}} & {\textbf{0.678 $\pm$ 0.242}} & {0.735 $\pm$ 0.111} & {\textbf{0.707}}\\
    {Modified nnUNet $^{**}$ \citep{myopschallenge-twostage-3}} & {0.672 $\pm$ 0.244} & {0.731 $\pm$ 0.109} & {0.702}\\
    {Modified nnUNet \citep{myopschallenge-twostage-3}} & {0.645 $\pm$ 0.236} & {0.690 $\pm$ 0.128} & {0.668}\\
    {EfficientSeg $^{**}$ \citep{EfficientSeg}} & {0.647 $\pm$ 0.279} & {0.709 $\pm$ 0.122} & {0.678}\\
    {CMS-UNet \citep{CMS-UNet}} & {0.581 $\pm$ 0.268} & {0.725 $\pm$ 0.110} & {0.653}\\
    {MF\&DFA-Net \citep{Dual-Path}} & {0.605 $\pm$ 0.263} & {0.656 $\pm$ 0.138} & {0.631}\\
     \midrule
    {MyoPS-Net-L (Ours)} & {0.647 $\pm$ 0.258} & {0.722 $\pm$ 0.135} & {0.685}\\
    {MyoPS-Net-L $^{**}$ (Ours)} & {0.661 $\pm$ 0.255} & {\textbf{0.742 $\pm$ 0.124}} & {0.702}\\
     \toprule
  \end{tabular}}
  \end{center}
 \end{table}

\section{Discussion and conclusion} %%%%% 5 Discussion and conclusion %%%%%
\label{section5}
In this work, we have proposed an end-to-end unified framework for MyoPS combining multi-sequence CMR images. 
Three major methodological contributions have been introduced.
Firstly, an effective cross-modal feature fusion architecture is proposed to extract instructive features from multi-sequence CMR images. 
Secondly, we propose a myocardium prior and consistency module to efficiently localize the desired pathologies.
Besides, an inclusiveness loss is introduced to impose constraints on pathology inclusiveness, which illustrates that scars are the subset of edema.
Experimental results proved the effectiveness of the proposed framework and demonstrated that this framework can be applied to four different scenarios in practical clinics.

Specially, the proposed framework chiefly concentrates on studying flexible combinations of multi-sequence CMR images for MyoPS. Ideally, employed full five-sequence CMR images will achieve the best scar and edema segmentation performance. For patients who are suitable for gadolinium contrast agents, taking LGE CMR is a guarantee for preciser diagnosis and further treatment of myocardial infarction. For those reluctant to inject contrast agents, \textit{i.e.}, healthy individuals undergoing a physical examination, mapping CMR can be an acceptable substitute for effective diagnosis of myocardial infarction.

In the above three conditions, we considered that either LGE CMR or mapping CMR is missing. Certainly, the proposed framework can also be applied to more complex conditions where more CMR sequences are missing. For example, we conducted an experiment only with bSSFP C0 and LGE CMR images, and the average Dice scores were 0.583 $\pm$ 0.153 and 0.628 $\pm$ 0.125 for myocardial scars and edema, respectively. Compared with MyoPS-Net-L in Table~\ref{proposed}, this model achieved comparative performance on scar segmentation, but an evidently lower Dice score on edema segmentation because of lacking T2 CMR images. However, segmenting edema with the combination of bSSFP C0 and LGE CMR images could be meaningless and misleading in clinical practice. This is because from clinical view, such combination does not provide pathological information to edema and the major proportion of edema region, which are scars, can be segmented from LGE CMR. Hence, the AI model may simply segment a marginally dilated region of the scars, to achieve high Dice scores, even though it cannot segment small isolated edema regions that do not contain scars. Therefore, we did not consider such combinations that lack clinical meaning or pathological sense.

Besides, the proposed framework can be applied to more than five CMR sequences. When there are more CMR sequences available, we can first analyze the clinical characteristics of these sequences. If these sequence shares similarities with LGE CMR or T2 CMR, we can then employ these sequences to perform scar or edema segmentation correspondingly. If these sequences resemble mapping CMR, we should perform ablation study like Section~\ref{4.3} and Table~\ref{Paramter} to determine the most suitable architecture.

There also exists two main limitations in this work. 
Firstly, we mainly focus on achieving MyoPS based on pre-aligned multi-sequence images using the MvMM method \citep{zhuang2016multivariate,zhuang2019multivariate}. 
However, these images are commonly not registered in clinical practice.
Therefore, a unified registration and segmentation framework of MS-CMR for MyoPS is valuable for clinical applications but still challenging due to cross-modality variances, and thus will be our future work.

Besides, one can see that C0, LGE and T2 CMR images can provide information for the myocardium, myocardial scars and edema, respectively.
Mapping CMR is expected to replace LGE CMR under certain conditions \citep{cine1-circulation2021}, but the proposed framework can only extract limited information from these sequences due to its poor quality and blurred pathology areas which is the second limitation.
In the future, we could employ cine CMR to facilitate the feature mining of mapping CMR. 
For example, \citet{cine2-media2020,cine4-media2018,cine3-media2020} developed the usage of cine CMR images on MyoPS with generative adversarial networks. 
\citet{cine1-circulation2021} proposed to synthesize an artificial CMR sequence which resembles LGE CMR via cine CMR images and T1 mapping CMR images. 
Therefore, there still exists significant space to make explorations on mapping CMR images with the auxiliary of cine CMR images.

\section*{Acknowledgments}
This work was funded by the National Natural Science Foundation of China (grant No. 61971142, 62111530195 and 62011540404) and the development fund for Shanghai talents (No. 2020015). 

\bibliographystyle{model2-names.bst}\biboptions{authoryear}
\bibliography{refs}

\end{document}